\documentclass[manuscript]{aastex}

\usepackage{xcolor}
\usepackage{graphicx}

\shorttitle{Kinetic Alfv\'en wave generation}
\shortauthors{V\'asconez et al.}

\begin{document}

\title{Kinetic Alfv\'en waves generation by large-scale phase-mixing}

\author{C. L. V\'asconez$^{1,2}$, F. Pucci$^1$, F. Valentini$^1$, S. Servidio$^1$, W. H. Matthaeus$^3$, F. Malara$^1$}
\affil{$^1$Dipartimento di Fisica, Universit\`a della Calabria,
    87036, Rende (CS), Italy.}
\affil{$^2$Observatorio Astron\'omico de Quito, Escuela Polit\'ecnica Nacional, Quito, Ecuador.}
\affil{$^3$Department of Physics and Astronomy, University of Delaware, DE 19716, USA.}

\input epsf 
\begin{abstract}
One view of the 
solar-wind turbulence is that the observed highly anisotropic 
fluctuations at spatial scales near the proton 
inertial length $d_p$ may be considered as 
Kinetic Alfv\'en waves (KAWs).
In the present paper, we show how 
phase-mixing of large-scale parallel propagating Alfv\'en waves is an efficient 
mechanism for the production of KAWs at wavelengths close to $d_p$ and at large propagation angle
with respect to the magnetic field. Magnetohydrodynamic (MHD), 
Hall-Magnetohydrodynamic (HMHD), and hybrid Vlasov-Maxwell (HVM) simulations modeling the 
propagation of Alfv\'en waves in inhomogeneous plasmas are performed. 
In linear regime, the role of dispersive effects 
is singled out by comparing MHD and HMHD results. Fluctuations produced by 
phase-mixing are identified as KAWs through a comparison of polarization of magnetic fluctuations and 
wave group velocity with analytical linear predictions. In the nonlinear regime, comparison of 
HMHD and HVM simulations allows to point out the role of kinetic effects 
in shaping the proton distribution function. We 
observe generation of temperature anisotropy with respect to the local magnetic field 
and production of field-aligned beams. 
The regions where the proton distribution function highly departs from thermal equilibrium are located inside 
the shear layers, where the KAWs are excited, this suggesting that the distortions of the proton distribution 
are driven by a resonant interaction of protons with KAW fluctuations. 
Our results are relevant in configurations where magnetic field inhomogeneities are present, as, for example, in the solar 
corona where the presence of Alfv\'en waves has been ascertained.
\end{abstract}



\section{Introduction}
Turbulence in plasmas is a complex phenomenon which is characterized 
by different regimes in different ranges of spatial and temporal 
scales. Turbulence in the solar wind has been extensively studied, 
both by detailed analyses of in-situ measurements 
and from a theoretical point of view; see \citet{bruno05} for a review. 
Such studies often adopt complementary views that 
the turbulence may be described either as collection of wave that interact nonlinearly, so-called wave-turbulence, or else as a collection of broad band, 
essentially zero frequency eddies or flux tubes that form a hierarchy 
of coherent structures. These approaches have been extensively reviewed
\citep{Barnes79,MatthaeusEA15} and we do not attempt 
a critical comparison in the present work. 
Instead, we adopt mainly a wave taxonomy of the fluctuations, 
based on linear theory in order to address a specific set of questions.
As motivation, we note that a variety of observations in the solar wind
\citep{bale05,sahraoui09} have suggested that 
fluctuations near the end of the magnetohydrodynamics 
inertial cascade range, and approaching the kinetic plasma range, may
consist primarily of Kinetic Alfv\'en waves (KAWs).   
Here we address in particular the nature of fluctuations produced
due to nonlinear interactions near the 
proton inertial length $d_p$ and investigate in 
some detail the basis for identifying them 
as KAWs.
We show how phase-mixing of large-scale parallel propagating Alfv\'en waves is an efficient mechanism for the production of KAWs 
at wavelengths close to $d_p$ and at large propagation angle with respect to the magnetic field. 
To support the interpretation as KAWs, 
we perform and analyze 
Magnetohydrodynamic (MHD), Hall-Magnetohydrodynamic (HMHD), and hybrid Vlasov-Maxwell (HVM) simulations that 
model the propagation of Alfv\'en waves and their fully nonlinear 
interaction with a nonuniform plasma background.  
We will be able to characterize 
fluctuations produced by this ``phase-mixing-''like 
interaction as highly oblique KAWs. 

At frequencies 
much lower than the proton gyrofrequency, solar wind turbulence is
dominated by Alfv\'enic fluctuations, which are characterized by 
highly-correlated velocity and magnetic fields and 
by low-level (with respect to the background values) density and 
magnetic field intensity variations \citep{belch71}. Moreover, in recent years several indications 
have been found of the presence of Alfv\'en waves also in the solar corona 
\citep{tomc07,tomc09} from where the solar wind emanates.
In the solar wind, Alfv\'enic fluctuations extend over a wide 
range of spatial scales, from large scales within the 
MHD range, down to lengths comparable or smaller than 
the proton inertial length $d_p$, where also finite Larmor radius effects become 
relevant. Moreover, in this scenario it is important to take into account the 
role of spectral anisotropy.
Theoretical studies, indeed, have shown 
that in a MHD turbulence the energy cascade preferentially takes place 
perpendicularly to the background magnetic field ${\bf B}_0$ (e.g., 
\citet{shebalin83,carbone90,OughtonEA94}).
Then, it is expected that at smaller scales the fluctuation energy tends to concentrate in 
nearly perpendicular wavevectors.
This idea is supported by observational data 
showing that the distribution of wavevectors of magnetic fluctuations
has a significant population quasi-perpendicular to the ambient magnetic field 
\citep{matthaeus86,matthaeus90}.
All the above effects, such as turbulence in a plasma with several characteristic lengthscales, 
Alfv\'enic correlations and spectral anisotropy, 
can interact with large scale structures such as shears and magnetic equilibria, 
with a subsequent generation of wave-like activity. The full picture need to be addressed 
with plasma simplified models, as well as the Vlasov-Maxwell treatment.

Waves belonging to the Alfv\'en branch, with wavelengths comparable 
with the proton inertial length $d_p$ and wavevectors nearly perpendicular to 
the mean magnetic field  ${\bf B}_0$ are often indicated as ``kinetic Alfv\'en waves''. 
During the last decades, KAWs have received considerable attention and 
have been studied in detail due to their possible role in 
a wave description of the turbulent cascade. 
Since the MHD cascade favors nearly perpendicular wavevectors, 
the expectation within a wave perspective
would be that KAWs are naturally present at scales 
of the order of $d_p$.
An extensive analysis of KAW physics is 
found in \citet{hollweg99} (see
also references therein for a more complete view on the subject). 
Many solar wind observational analyses
\citep{bale05,sahraoui09,podesta12,salem12,chen13,kiyani13}, theoretical works 
\citep{howes08a,scheko09,sahraoui12} as well as 
numerical simulations 
\citep{gary04,howes08b,tenbarge12} have suggested 
that KAWs can play 
an important role in the dissipation of turbulent energy. 
Due to a nonvanishing electric field parallel 
component associated with KAWs, these waves have also been considered 
in the problem of particle acceleration \citep{voitenko04,decamp06}. 
Recently, 
\citet{vasconez14} have studied collisionless Landau damping and 
wave-particle resonant interactions in KAWs.

There are also simplified problems, less complex than 
fully developed turbulence, 
in which one finds the formation of small scales in the 
direction perpendicular to an applied magnetic field ${\bf B}_0$. 
It is well known that this effect appears 
in the context of MHD when 
imposed parallel propagating waves interact
with an inhomogeneous background 
consisting either of pressure balanced structures or velocity shears
\citep{GhoshEA98}.
In 2D equilibria, where the Alfv\'en velocity
varies in directions perpendicular to the magnetic field, two
mechanisms have been investigated in detail: (1) phase-mixing
\citep{heyvaerts83}, in which differences in group velocity 
at different locations progressively bend wavefronts; and
(2) resonant absorption which concentrates the wave energy in a
narrow layer where the wave frequency locally matches a characteristic 
frequency (Alfv\'en or cusp). These processes have been
studied both by investigating normal modes of the inhomogeneous structure 
\citep{kappraff77,mok85,steinolfson85,davila87,hollweg87,califano90,califano92}
and by considering the evolution of an initial 
disturbance \citep{lee86,malara92,malara96}.
Effects of density stratification and magnetic line divergence \citep{ruderman98}, 
as well as nonlinear coupling with compressive 
modes \citep{nakariakov97,nakariakov98}, and  
evolution of localized pulses \citep{tsiklauri02,tsiklauri03} have been considered. 
The propagation of MHD waves in inhomogeneous magnetic fields
containing null points has also been studied in detail 
(\citet{landi05}; see also \citet{mclaughlin10} for a review). 
Phase-mixing in 3D inhomogeneous equilibria 
has also been considered in the small wavelength limit \citep{similon89} using a WKB approximation 
\citep{petkaki98,malara00}, also within the problem of coronal heating 
\citep{malara03,malara05,malara07}. Particle acceleration in phase-mixing 
of Alfv\'en waves in a dispersive regime has been studied by \citet{tsiklauri05} using 
particle-in-cell simulations, both in 2D \citep{tsiklauri11} and in 3D \citep{tsiklauri12} 
configurations. Finally, instabilities generating KAWs in plasma 
with transverse density modulations have been considered by \citet{wu13}.
Similar ideas involving dissipative mechanisms related to interaction of 
Alfv\'en waves or KAWs and phase-mixing 
have been examined in the context of the magnetospheric plasma sheet
\citep{LysakSong11} and in coronal loops \citep{OfmanAschwanden02}.

The above considerations suggest that phase-mixing of Alfv\'en waves 
might represent a mechanism to produce KAWs, when  
the wavelength of waves becomes enough small to be comparable 
with the proton inertial length $d_p$. This effect could work, for 
instance, in the solar corona where the background magnetic field is clearly
inhomogeneous and where the presence of Alfv\'en waves has been 
ascertained. This represents an indication about the nature of small-scale 
fluctuations that could be present in the coronal plasma.
The aim of the present paper is to investigate directly 
the generation of KAW fluctuations
associated with Alfv\'en wave phase-mixing. This study will be 
performed numerically by using both a HMHD code and a 
Vlasov-hybrid code. The former include proton skin depth effects, 
while the latter allows identification of kinetic effects 
such as 
wave-particle resonances and temperature anisotropy. 
In Section 2 the model will be presented 
along with a linear analysis of wave properties;
the results derived by the HMHD in the linear regime are described in Section 3, while 
the nonlinear regime and the results of the 
Vlasov-hybrid code will be described in Section 4; 
a discussion and a summary of results are given in Section 5.

\section{Hall-MHD model}
Consider a fluid plasma composed of protons and electrons. For 
phenomena at sufficiently large scales the electron mass can be neglected 
$m_e \ll m_p$ and the quasi-neutrality condition holds: $n_e \simeq n_p$, 
with $n_e$ and $n_p$ the number densities of electrons and protons, 
respectively. Moreover, we assume that the temperatures of protons and 
electrons are equal $T_e=T_p=T$. Finally, the displacement current term 
is neglected in the Ampere equation. In these conditions the 
plasma dynamics can be described by the 
HMHD equations:
\begin{equation}\label{HMHD1}
\frac{\partial \rho}{\partial t}+\nabla \cdot (\rho {\bf v})=0
\end{equation}
\begin{equation}\label{HMHD2}
\frac{\partial {\bf v}}{\partial t} +({\bf v}\cdot \nabla){\bf v}=
-\frac{{\tilde \beta}}{2\rho}\nabla (\rho T)+
\frac{1}{\rho} \left[(\nabla \times {\bf B})\times {\bf B}\right]
\end{equation}
\begin{equation}\label{HMHD3}
\frac{\partial {\bf B}}{\partial t}=
\nabla \times \left[ {\bf v}\times {\bf B} -
\frac{{\tilde \epsilon}}{\rho} \left(\nabla \times {\bf B}\right)\times {\bf B} \right]
\end{equation}
\begin{equation}\label{HMHD4}
\frac{\partial T}{\partial t} + ({\bf v}\cdot \nabla)T +
(\gamma -1)T(\nabla \cdot {\bf v})=0
\end{equation}
Equations (\ref{HMHD1})-(\ref{HMHD4}) contain only dimensionless 
quantities. In particular,  
mass density $\rho$ (which is only due to protons) is normalized
to a typical density 
${\tilde \rho}$, temperature $T$ to a typical value ${\tilde T}$, 
the pressure $p=\rho T$ due to both protons and electrons 
is normalized to the value ${\tilde p}=2\kappa_B {\tilde \rho}{\tilde T}/m_p$, 
with $\kappa_B$ the Boltzmann constant and $m_p$ the proton mass. 
The spatial coordinates are normalized to a typical 
length ${\tilde L}$, magnetic field ${\bf B}$ is normalized to a typical 
magnetic field ${\tilde B}$, 
fluid velocity ${\bf v}$ to the typical Alfv\'en speed 
${\tilde c}_A={\tilde B}/(4\pi {\tilde \rho})^{1/2}$, 
time $t$ to the Alfv\'en time ${\tilde t}_A={\tilde L}/{\tilde c}_A$. 
Finally, the plasma beta ${\tilde \beta}={\tilde p}/({\tilde B}^2/8\pi)$ 
is a typical value for the kinetic to magnetic pressure ratio; 
$\gamma =5/3$ is the adiabatic index; 
${\tilde \epsilon} = {\tilde d}_p/{\tilde L} \ll 1$ is the Hall parameter measuring the 
relative amplitude of the Hall term with respect to the 
${\bf v}\times {\bf B}$ term in the Ohm's law, 
${\tilde d}_p={\tilde c}_A/{\tilde \Omega}_{cp}={\tilde c}_A m_p c/(q{\tilde B})$ 
being the proton inertial length. Equations 
(\ref{HMHD1})-(\ref{HMHD4}) reduce to the ordinary single fluid compressible 
MHD equations in the 
limit ${\tilde \epsilon} \rightarrow 0$.
The equations are ideal and adiabatic due to the absence of 
viscosity in Eq. (2), resistivity in Eq. (3) and heat 
sources and heat conduction in Eq. (4).
For the simulations carried out below these effects are 
present in some form (not necessarily in the form of a 
fluid model); however at present they are omitted for clarity. 

\subsection{Wave properties}
We consider first a uniform equilibrium state characterized by 
homogeneous dimensionless density $\rho_0$, temperature $T_0$, magnetic 
field ${\bf B}_0$ and vanishing velocity ${\bf v}_0=0$. 
Waves propagating in the above equilibrium can be studied 
as usual: a small amplitude perturbation is superposed 
on the equilibrium; equations (\ref{HMHD1})-(\ref{HMHD4}) are 
linearized with respect to the perturbation amplitude and 
perturbation fields are Fourier transformed both in space and time. 
Imposing nonvanishing perturbations leads to the following 
dispersion relation:
\begin{equation}\label{disprel}
\omega^6+C_1 \omega^4+C_2 \omega^2+C_3=0
\end{equation}
with
\begin{equation}\label{C1}
C_1=
-k_{||}^2 \left( 2c_{A0}^2+c_{s0}^2 \right)
-k_\perp^2 \left( c_{A0}^2+c_{s0}^2 \right)
-\frac{c_{A0}^4 k_{||}^2 \left( k_{||}^2 + k_\perp^2\right)}{\omega_{p0}^2}
\end{equation}
\begin{equation}\label{C2}
C_2=
c_{A0}^2 k_{||}^2 \left(c_{A0}^2+2c_{s0}^2 \right)
\left( k_{||}^2 + k_\perp^2 \right)+
\frac{c_{A0}^4 c_{s0}^2 k_{||}^2
\left( k_{||}^4 +2 k_{||}^2 k_\perp^2 + k_\perp^4 \right)}{\omega_{p0}^2}
\end{equation}
\begin{equation}\label{C3}
C_3=
-c_{A0}^4 c_{s0}^2 k_{||}^4 
\left( k_{||}^2+k_\perp^2 \right)
\end{equation}
In equations, (\ref{disprel})-(\ref{C3}) $\omega$ is the wave frequency normalized to 
${\tilde t}_A^{-1}$; 
$k_{||}$ and $k_\perp$ are the wavevector components parallel and perpendicular 
to ${\bf B}_0$, respectively, both normalized to ${\tilde L}^{-1}$; 
$c_{A0}=B_0/(4\pi \rho_0)^{1/2}$ and 
$c_{s0}=({\tilde \beta} \gamma T_0/2)^{1/2}$ are the Alfv\'en speed 
and the sound speed associated with the equilibrium, respectively; 
$\omega_{p0}=(q {\tilde B}B_0/(m_p c))({\tilde L}/{\tilde c}_A)$ is the  
normalized proton gyrofrequency. 
The squared sound speed to Alfv\'en speed ratio will be 
indicated by $\beta_0=c_{s0}^2/c_{A0}^2$.

The expressions (\ref{disprel})-(\ref{C3}) are equivalent to those 
found in a two-fluid model by \citet{vasconez14} if the electron inertia is neglected. 
Equation (\ref{disprel}) has been analytically solved \citep{vasconez14} 
using the Vieta's substitution method \citep{birkhoff77} 
for the calculation of the complex roots of a third-degree algebraic equation. 
For given values of $k_{||}$ and $k_\perp$ the three solutions found for 
$\omega^2$ are real and positive and correspond to the Alfv\'en, fast magnetosonic (FM) 
and slow magnetosonic (SM) branches, respectively.
In particular, KAWs correspond to the Alfv\'en branch for 
$k_{||} \ll k_\perp \sim \omega_{p0}/c_{A0}$.

From the linear analysis the expressions for 
the amplitudes of fluctuations can be calculated. We use a reference frame 
where the $x$-axis is along ${\bf B}_0$ while the wavevector 
${\bf k}$ is in the $xy$ plane. The perturbation components 
for a wave with wavevector ${\bf k}=k_{||}{\bf e}_x+k_\perp{\bf e}_y$ 
can be expressed in the following form:
\begin{equation}\label{v1x}
v_{1x}=a \frac{c_{A0}^2 c_{s0}^2 k_{||}^2 k_\perp \omega_{p0}}
{\omega^2 \left( \omega^2 - c_{s0}^2 k^2\right)}
\left(1-\frac{\omega^2}{c_{A0}^2 k_{||}^2}\right) \sin(k_{||}x+k_\perp y-\omega t+\phi)
\end{equation}
\begin{equation}\label{v1y}
v_{1y}=a \frac{c_{A0}^2 k_{||} \omega_{p0} \left( \omega^2 - c_{s0}^2 k_{||}^2 \right)}
{\omega^2 \left( \omega^2 - c_{s0}^2 k^2\right)}
\left(1-\frac{\omega^2}{c_{A0}^2 k_{||}^2}\right) \sin(k_{||}x+k_\perp y-\omega t+\phi)
\end{equation}
\begin{equation}\label{v1z}
v_{1z}=-a \frac{c_{A0}^2 k_{||}}{\omega} \cos(k_{||}x+k_\perp y-\omega t+\phi)
\end{equation}
\begin{equation}\label{B1x}
B_{1x}=a B_0\frac{k_{||} k_\perp \omega_{p0}}{\omega k^2}
\left(1-\frac{\omega^2}{c_{A0}^2 k_{||}^2}\right) \sin(k_{||}x+k_\perp y-\omega t+\phi)
\end{equation}
\begin{equation}\label{B1y}
B_{1y}=-a B_0 \frac{k_{||}^2\omega_{p0}}{\omega k^2}
\left(1-\frac{\omega^2}{c_{A0}^2 k_{||}^2}\right) \sin(k_{||}x+k_\perp y-\omega t+\phi)
\end{equation}
\begin{equation}\label{B1z}
B_{1z}=a B_0\cos(k_{||}x+k_\perp y-\omega t+\phi)
\end{equation}
where $a$ is the perturbation amplitude, $\phi \in \lbrack 0, 2\pi \rbrack$ 
is the phase and $\omega=\omega({\bf k})$ is derived 
from the dispersion relation (\ref{disprel}).

In what follows it will be useful to have an expression for the component of the wave group velocity 
perpendicular to ${\bf B}_0$, $v_{g\perp}=\partial \omega/\partial k_\perp$. 
This has been calculated in the following way: 
we indicate by $F(\omega,k_{||},k_\perp)$ the LHS of equation (\ref{disprel})
and by $\omega(k_{||},k_\perp)$ a solution of equation (\ref{disprel}).
Substituting $\omega(k_{||},k_\perp)$ in the place of $\omega$ into 
equation (\ref{disprel}) we obtain an identity
\begin{equation}\label{ident}
F(\omega(k_{||},k_\perp),k_{||},k_\perp)\equiv 0
\end{equation}
which holds for any value of $k_{||}$ and $k_\perp$. The derivative of 
equation (\ref{ident}) with respect to $k_\perp$ is 
\begin{displaymath}
\frac{d F}{d k_\perp}=\frac{\partial F}{\partial \omega}
\frac{\partial \omega}{\partial k_\perp}+\frac{\partial F}{\partial k_\perp}=0
\end{displaymath}
from which we obtain
\begin{equation}\label{vgparperp}
v_{g\perp}=\frac{\partial \omega}{\partial k_\perp}=
-\frac{\partial F/\partial k_\perp} 
{\partial F/\partial \omega}
\end{equation}
The derivatives of $F$ appearing in the RHS of equation (\ref{vgparperp}) can 
be calculated using equations (\ref{disprel})-(\ref{C3}); their 
expressions are: 
\begin{eqnarray}\label{dFdkperp}
\frac{\partial F}{\partial k_\perp}&=&
-2\left[ k_\perp \left(c_{A0}^2+c_{s0}^2 \right)+
\frac{k_\perp k_{||}^2 c_{A0}^4}{\omega_{p0}^2}\right] \omega^4 + \nonumber \\
&& 2\left[ k_\perp k_{||}^2 \left(2 c_{s0}^2 c_{A0}^2 + c_{A0}^4 \right) +
2 \frac{\left( k_\perp k_{||}^4 + k_\perp^3 k_{||}^2 \right) c_{A0}^4 c_{s0}^4}{\omega_{p0}^2} \right] \omega^2
-2 k_\perp k_{||}^4 c_{s0}^2 c_{A0}^4
\end{eqnarray}
\begin{eqnarray}\label{dFdomega}
\frac{\partial F}{\partial \omega} = 
6 \omega^5 -4 \left[ k^2 \left(c_{A0}^2 + c_{s0}^2 \right) + k_{||}^2 c_{A0}^2 
+\frac{k^2 k_{||}^2 c_{A0}^2}{\omega_{p0}^4} \right]\omega^3 + \nonumber \\
2 \left[ k^2 k_{||}^2 \left( 2 c_{s0}^2 c_{A0}^2 + c_{a0}^4 \right) +
\frac{k_{||}^2 k^4 c_{A0}^4 c_{s0}^2}{\omega_{p0}^2} \right] \omega
\end{eqnarray}
where $k^2=k_{||}^2+k_\perp^2$. Since phase-mixing 
increases $k_\perp$ leaving $k_{||}$ constant (as shown in the 
following section), in Fig. \ref{Fig:gvperp} 
the perpendicular group velocity $v_{g\perp}$ of the Alfv\'en, FM and SM 
modes is plotted as a function of $k_\perp$ in a range $k_\perp \ge k_{||}$, for 
$k_{||}=1$ and for $\beta_0=1.25$ and $\beta_0=2.08$. The choice of these two values of $\beta_0$ 
will be justified in the next section. 
From Fig. \ref{Fig:gvperp} we see that $v_{g\perp}$ of the three modes increases with increasing $k_\perp$,
becoming nearly constant at large propagation angles. For $k_\perp \ge k_{||}=1$, $v_{g\perp}$ 
is positive for Alfv\'en and FM waves while is negative for SM waves. Moreover, $v_{g\perp}$ of 
FM waves is much larger than that of the Alfv\'en branch; this is consistent with 
a quasi-isotropic propagation velocity of FM waves. The behavior of $v_{g\perp}$ for 
negative $k_\perp$ can be inferred from Fig. \ref{Fig:gvperp} taking into account that 
$v_{g\perp}(k_{||},-k_\perp)=-v_{g\perp}(k_{||},k_\perp)$.
These features will be used to identify the nature of fluctuations
generated by phase-mixing.

\section{MHD and Hall-MHD simulations of phase-mixing}

\begin{figure}
\plotone{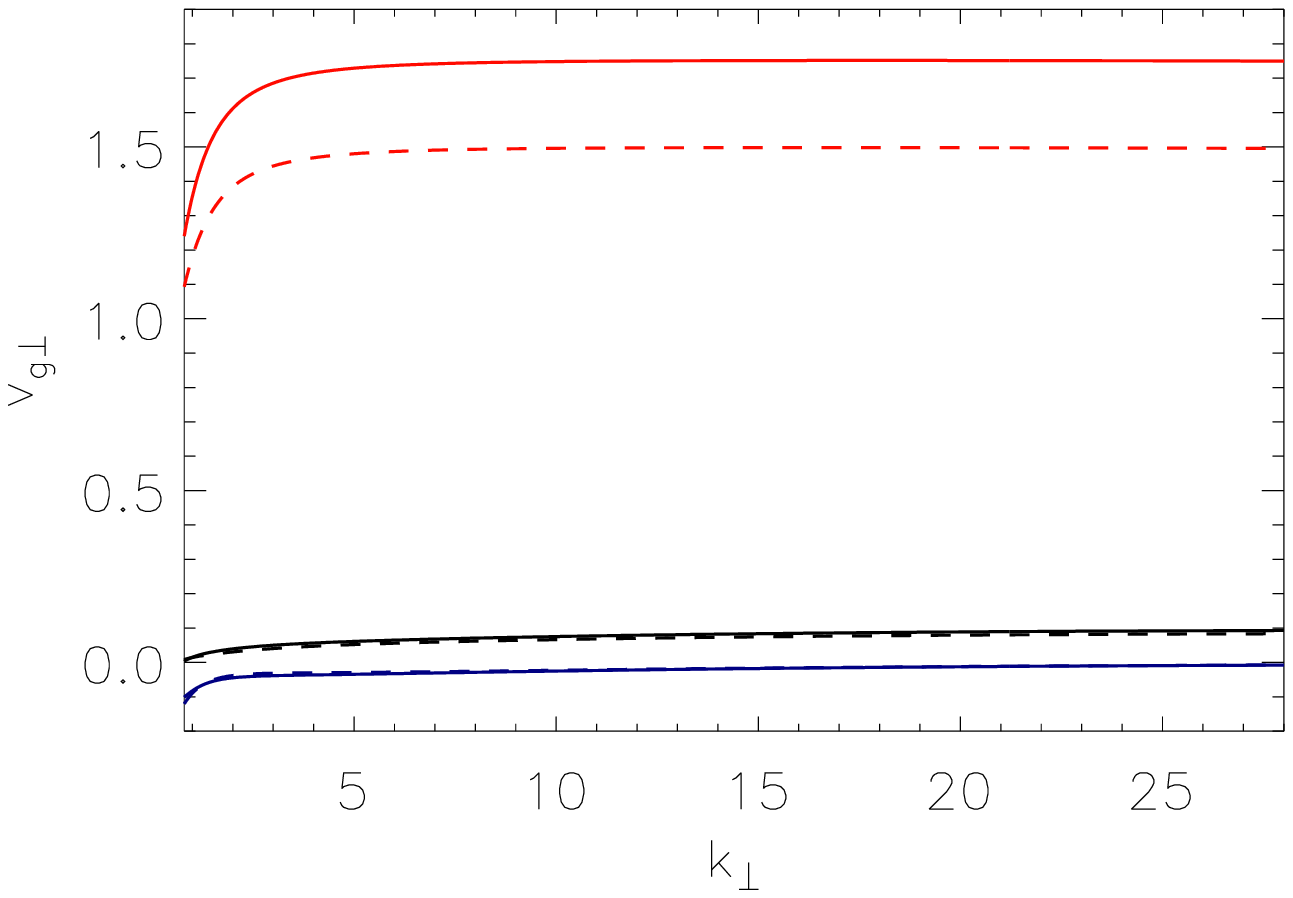}
\caption{The perpendicular group velocity $v_{g\perp}$ of the Alfv\'en (black lines), 
FM (red lines) and SM (blue lines) modes is plotted as a function 
of $k_\perp$, for $k_{||}=1$ and for $\beta_0=2.08$ (full lines) and $\beta_0=1.25$ (dashed lines).
\label{Fig:gvperp}}
\end{figure}

\begin{figure}
\plotone{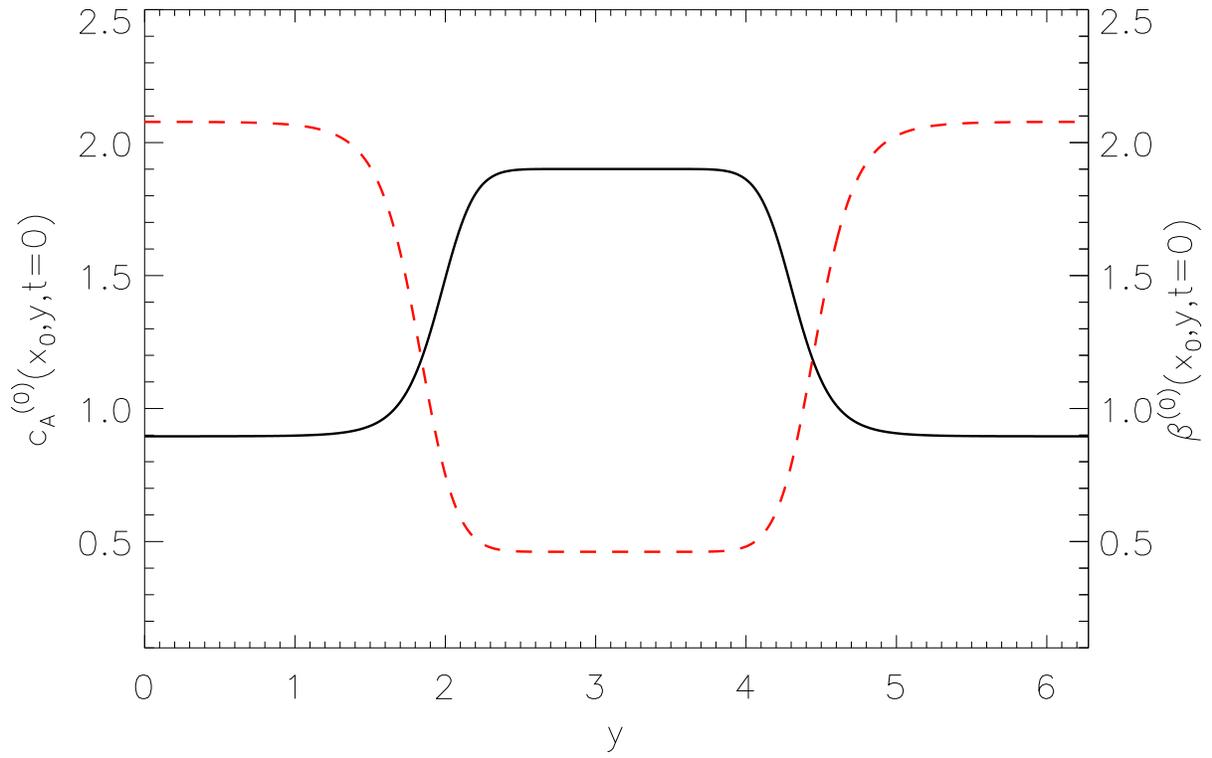}
\caption{The Alfv\'en velocity $c_A^{(0)}$ (black solid line) and $\beta^{(0)}$ (red dashed line) 
associated with the equilibrium structure are plotted as functions of the $y$ coordinate.\label{Fig:alfbeta}}
\end{figure}

\begin{figure}
\plotone{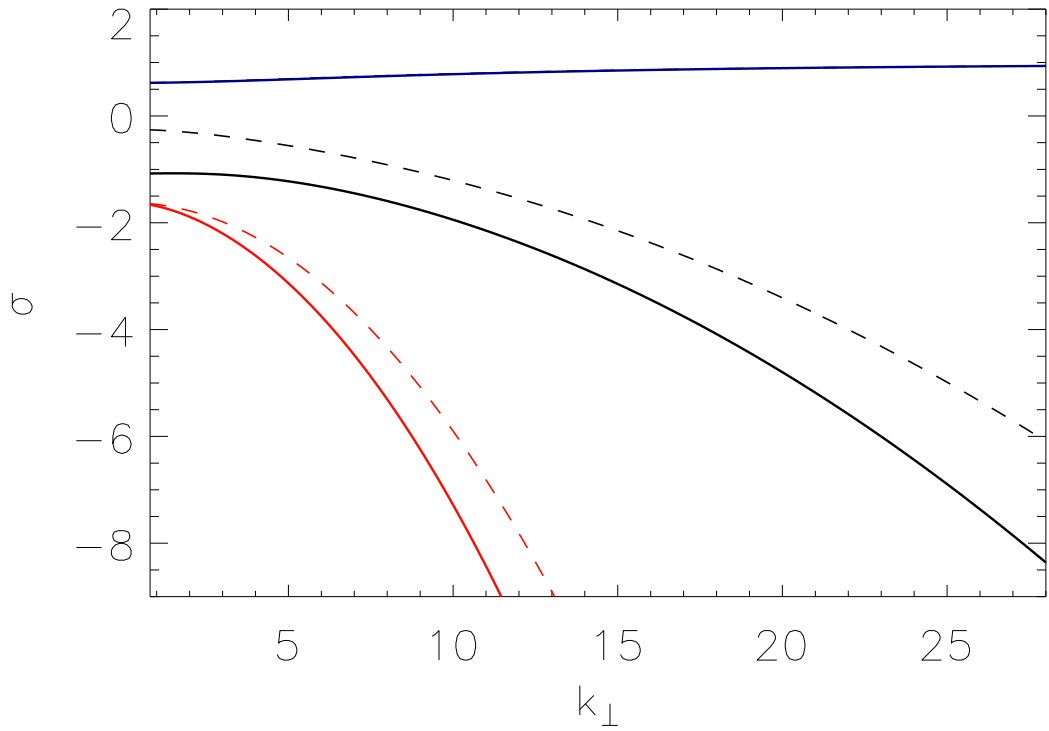}
\caption{The quantity $\sigma$ is plotted as a function of $k_\perp$, for $k_{||}=1$ and 
for $\beta_0=2.08$ (full lines) and $\beta_0=1.25$ (dashed lines). Black lines correspond to 
the Alfv\'en mode, red lines to the FM mode, and blue lines to the SM mode. In the latter case
the curves corresponding to the two values of $\beta_0$ are superposed.
\label{Fig:polar}}
\end{figure}

We consider a 2.5D configuration where all the 
physical quantities depend only on two spatial variables ($x$ and $y$), 
but vector quantities can have three nonvanishing components. 
The above quantities are defined in a spatial domain $D=\lbrace (x,y) \rbrace = 
\lbrack 0, 2\pi \rbrack \times \lbrack 0, 2\pi \rbrack$ 
(in dimensionless units), where 
periodic boundary conditions are imposed both in the $x$ and $y$ directions. 
We consider a nonuniform equilibrium structure in 
$D$, where physical quantities vary only along the $y$-direction, or are uniform.
Quantities relative to the equilibrium are indicated by the upper index "(0)". 
The equilibrium magnetic field is
\begin{equation}\label{B0}
{\bf B}^{(0)}=B^{(0)}(y) {\bf e}_x
\end{equation}
where
\begin{equation}\label{B0s}
B^{(0)}(y)=1+\frac{b_m -1}{1+\left( \displaystyle{\frac{y-\pi}{2\pi h}}\right)^r}
+\alpha \left( \frac{y}{\pi}-1 \right)^2
\end{equation}
and ${\bf e}_x$ is the unit vector along $x$.
The dimensionless parameters in the expression (\ref{B0s}) have the following values: 
$b_m=1.5$, $h=0.2$, $r=10$ and
\begin{equation}\label{alpha}
\alpha=\frac{(b_m-1)r}{2 (2h)^r \left[ 1+\left(\displaystyle{\frac{1}{2h}}\right)^r \right]^2}
\simeq 2.62 \times 10^{-4}
\end{equation}
The function $B^{(0)}(y)$ is symmetrical with respect to the central point $y=\pi$, 
where it reaches its maximum value $b_m$. The small term containing $\alpha$ 
in the expression (\ref{B0s}) has been added in order to have a vanishing 
first derivative of $B^{(0)}(y)$ at the two boundaries $y=0$ and $y=2\pi$. 
Both $B^{(0)}(y)$ and $dB^{(0)}/dy$ are periodic functions in the interval
$\lbrack 0,2\pi \rbrack$. However, higher order derivatives of $B^{(0)}(y)$ are 
not exactly periodic; as a consequence, the Fourier spectrum of $B^{(0)}(y)$ 
has a tail at high wavenumbers. To avoid this drawback, the expression 
(\ref{B0s}) has been corrected by filtering out harmonics with wavenumbers 
larger than $70$ in its spectrum. The filter does not sensibly alter 
the profile $B^{(0)}(y)$.
The equilibrium magnetic field is almost constant 
both in the central half and on the two lateral parts of the domain, while 
two sharp shear layers (current sheets) are located in between these 
uniform regions.
The equilibrium temperature, which is equal both for protons and electrons,
has been chosen as uniform, while the fluid velocity is identically 
zero:
\begin{equation}\label{T0v0}
T^{(0)}=1\;\; , \;\; {\bf v}^{(0)}=0
\end{equation}
The equilibrium mass density $\rho^{(0)}(y)$ is determined by total pressure 
equilibrium:
\begin{equation}\label{pequi}
\frac{{\tilde \beta}}{2}\rho^{(0)}(y) T^{(0)}+\frac{{B^{(0)}}^2(y)}{2}=P_T^{(0)}
\end{equation}
where $P_T^{(0)}=1.748$ and ${\tilde \beta}=2$.
The Alfv\'en and sound velocities associated with equilibrium structure are given by 
$c_A^{(0)}(y)=B^{(0)}(y)/\lbrack{\rho^{(0)}(y)}\rbrack^{1/2}$ and 
$c_S^{(0)}=({\tilde \beta} \gamma T^{(0)}/2)^{1/2}$, respectively. 
The local plasma $\beta$ is $\beta^{(0)}(y)=\lbrack{c_s^{(0)}/c_A^{(0)}(y)}\rbrack^2$. 
The profiles of $c_A^{(0)}(y)$ and $\beta^{(0)}(y)$ 
are shown in Fig. \ref{Fig:alfbeta}. The Alfv\'en velocity is larger in the lateral parts of 
$D$ than in the center; the inhomogeneity of $c_A^{(0)}$ is responsible for phase-mixing 
of Alfv\'en waves. We note that $\beta^{(0)} <1$ in the central region, while 
$\beta^{(0)}$ becomes larger than 1 when approaching the boundaries $y=0$ and $y=2\pi$. 
In particular, $\beta^{(0)}=1.25$ in the middle of the shear layers while $\beta^{(0)}=2.08$ 
in the lateral homogeneous regions; these two values have been used to calculate 
the profiles of Fig.s \ref{Fig:gvperp} and \ref{Fig:polar}. 

At the initial time an Alfv\'enic perturbation has been superposed on the above 
equilibrium. Quantities relative to the perturbation are indicated by the upper 
index "(1)". The initial magnetic field and velocity perturbation are given by 
\begin{equation}\label{B1v1}
{\bf B}^{(1)}(x,y,t=0)=a \cos (x) {\bf e}_z \;\;\; , \;\;\; 
{\bf v}^{(1)}(x,y,t=0)=-a \lbrack \rho^{(0)}(y) \rbrack^{-1/2} \cos (x) {\bf e}_z
\end{equation}
so that ${\bf v}^{(1)}=-(c_A^{(0)}/B^{(0)}){\bf B}^{(1)}$. 
The quantity $a$ gives the amplitude 
of the initial perturbation. Initial density and temperature 
fluctuations are vanishing: $\rho^{(1)}(x,y,t=0)=0$, $T^{(1)}(x,y,t=0)=0$.

The MHD case is described by equations (\ref{HMHD1})-(\ref{HMHD4}) with ${\tilde \epsilon}=0$. 
In this case and in the small-amplitude limit $a\ll 1$ the equations may be linearized and 
above initial conditions 
evolve in time according to the equations:
\begin{equation}\label{phasemix}
{\bf B}^{(1)}(x,y,t)=a \cos \left[ x-c_A^{(0)}(y)t\right] {\bf e}_z
\; , \; 
{\bf v}^{(1)}(x,y,t)=-a \lbrack \rho^{(0)}(y) \rbrack^{-1/2} 
\cos \left[ x-c_A^{(0)}(y)t\right] {\bf e}_z
\end{equation}
indicating that the initial perturbation propagates along 
${\bf B}^{(0)}$ at the local Alfv\'en speed. Thus, transverse variations 
of $c_A^{(0)}$ generates in the perturbation increasingly smaller scales 
which are localized within the shear layers. This phenomenon represents 
phase-mixing of an Alfv\'en wave \citep{heyvaerts83}.

The fully nonlinear
equations (\ref{HMHD1})-(\ref{HMHD4}) have been solved numerically
with the above-specified boundary and initial conditions. 
The HMHD numerical 
code employs a 2D Fourier pseudospectral method to calculate spatial derivatives 
and time integration is performed via a second-order Runge-Kutta scheme. Aliasing 
errors in the evaluation of nonlinear terms are partially removed by a 2/3 truncation in 
the spectral space. Equations have been solved using the scheme described in \citet{GhoshEA93}. 
Hyper-viscosity and hyper-resistivity terms 
(fourth-order derivatives) have been added in equations (\ref{HMHD2}) and 
(\ref{HMHD3}) in order to obtain numerical stability with dissipation 
concentrated only at the smallest spatial scales. 

Different runs have been performed; 
Table 1 summarizes 
the values of parameters used 
in the various cases, with $n_x$ and $n_y$ 
the number of gridpoints in the $x$ and $y$ direction.
RUN 1 corresponds to the MHD case (${\tilde \epsilon}=0$) with a low-amplitude 
perturbation ($a=0.01$). 
Though the analytical solution for an infinitesimal amplitude is known
(equations (\ref{phasemix})), we performed this run to single out differences of 
a purely MHD case with respect to a HMHD case (${\tilde \epsilon} \ne 0$). 
Fluctuating fields are defined as $\delta g=g-\langle g \rangle_x$, where $g$ 
represents a physical quantity from which the spatial average along $x$ has been subtracted 
in order to eliminate the 
contribution from the equilibrium structure. The effect of phase-mixing is visible 
in the time evolution of $\delta v_z$ and $\delta B_z$ (not shown): the wave propagates 
from left to right but with a velocity which is larger in the central part than in the lateral 
parts of the domain. As a consequence, within 
the two shear layers the wave profile is stretched  
with small-scale gradients generated 
in the transverse ($y$) direction.
The other components of $\delta {\bf v}$ and $\delta {\bf B}$, as well as 
the density and temperature fluctuations, have a much lower amplitude which is 
of the order of $10^{-4}$, or smaller. Moreover, these fluctuations have a wavelength 
in the $x$ direction which is half of the wavelength of $\delta v_z$ and $\delta B_z$. These two 
features clearly indicate that in the MHD case 
$\delta v_x$, $\delta v_y$, $\delta B_x$, $\delta B_y$, $\delta \rho$ 
and $\delta T$  
are generated by small nonlinear effects that
are quadratic in the Alfv\'en wave amplitude.

In RUN 2 we considered again a small-amplitude initial wave ($a=0.01$), but now 
dispersive effects have been switched on by setting the coefficient of the Hall term
${\tilde \epsilon}=0.125$. This corresponds to set the proton inertial length 
${\tilde d}_p=0.125$ (in normalized units). Thus, it is expected that dispersive effects 
are no longer negligible as soon as the perturbation 
wavevector $k$ has increased enough (in consequence of phase-mixing) 
to become comparable with ${\tilde d}_p^{-1}$. 
This condition is reached inside the shear layers. 
The time $t_d$ at which $k \sim {\tilde d}_p^{-1}$ can be estimated in the 
following way: during phase-mixing the transverse wavevector component $k_y$ increases 
in time according to the equation (e.g. \citet{petkaki98})
\begin{equation}\label{kypm}
k_y(t) \sim k_{y0} - \left(\frac{dc_{A0}}{dy}\right)_s k_x t
\end{equation}
where $(dc_{A0}/dy)_s$ is an estimation of the Alfv\'en velocity gradient in the 
shear layers, while $k_x$ is constant and $k_{y0}=k_y(t=0)$. In our case 
$k_x=1$, $k_{y0}=0$ and $(dc_{A0}/dy)_s \simeq -1$ (in the shear layer at $x\simeq 4.5$). 
At the time $t=t_d$ 
we have $k_y\gg k_x$; then, from the condition $k_y(t_d)d_p\simeq 1$ 
we get the estimate
\begin{equation}\label{td}
t_d \simeq -\frac{1}{d_p k_x \left(\displaystyle{\frac{dc_{A0}}{dy}}\right)_s} \simeq 8
\end{equation}
In our configuration it is expected that at times $t \gtrsim t_d$ phase-mixing could generate 
KAWs, i.e., perturbations belonging to the Alfv\'en branch with a 
quasi-perpendicular wavevector ($k_y\gg k_x$). 
We now explore this possibility in some detail. 

\begin{figure}
\plotone{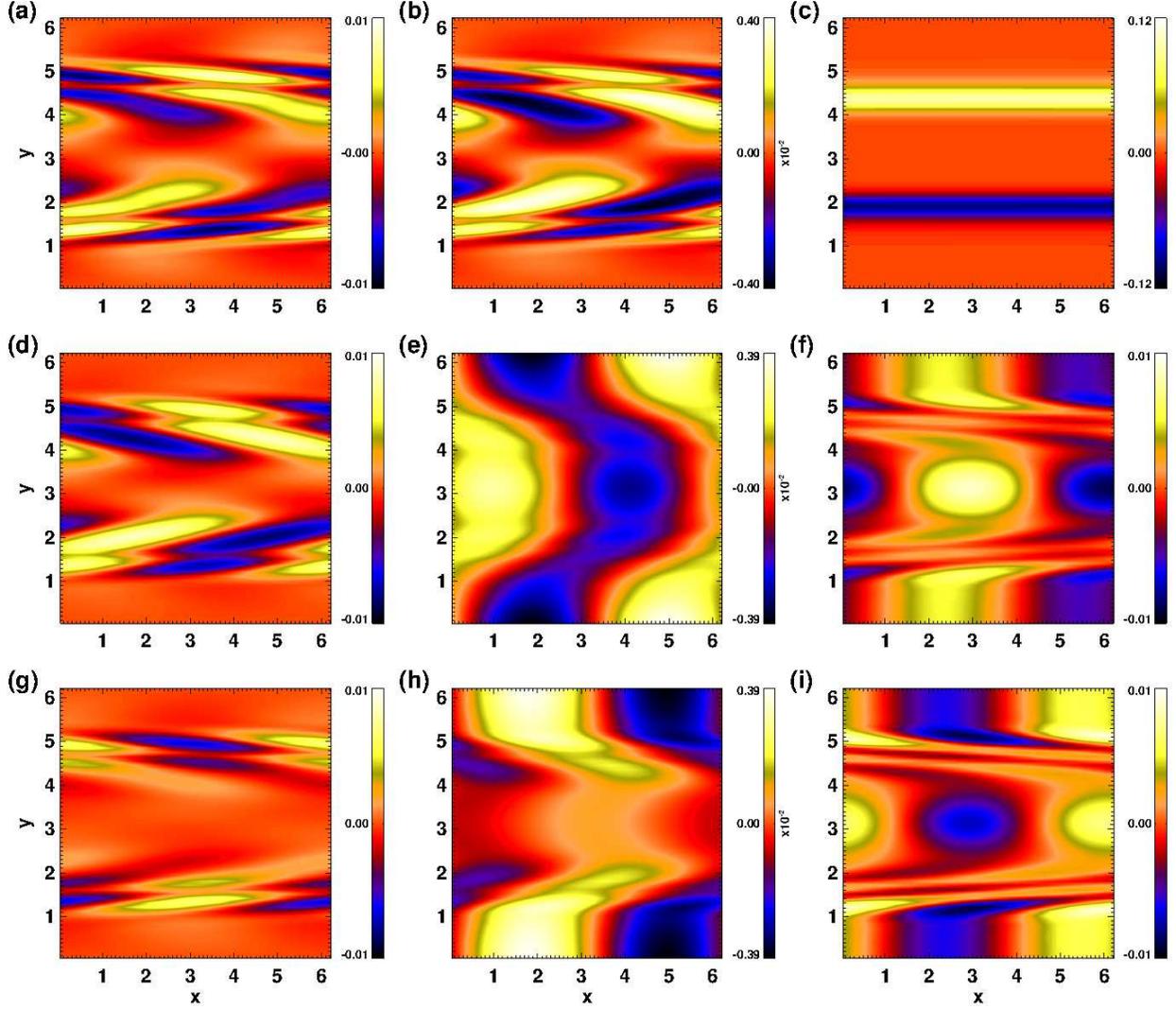}
\caption{$\delta \rho$ (a), $\delta T$ (b), $j_z$ (c), $\delta v_x$ (d), $\delta v_y$ (e),
$\delta v_z$ (f), $\delta B_x$ (g), $\delta B_y$ (h), $\delta B_z$ (i), plotted as functions of $x$ and
$y$, at the time $t = 13.1$ for the low-amplitude HMHD run (RUN 2). \label{Fig:HMHDlin}}
\end{figure}

In Fig. \ref{Fig:HMHDlin} the fluctuating fields $\delta \rho$, $\delta T$, 
$\delta {\bf v}$ and $\delta {\bf B}$ are represented at the time $t=13.1$, 
along with the current density component $j_z$ which has been plotted in order to 
localize the equilibrium shear layers. The time $t=13.1$ is larger than $t_d$, thus we 
expect to observe effects due to finite ion inertial length. 
We observe that $\delta v_x$, $\delta B_x$, $\delta \rho$ and $\delta T$ are now of the same order as 
$\delta v_z$ and $\delta B_z$ and the parallel wavelength is $2\pi$ for all these fields. 
This indicates that in the HMHD case $\delta v_x$, $\delta B_x$, $\delta \rho$ and $\delta T$ fluctuations are 
not due to nonlinear effects but they are 
part of the same perturbation as $\delta v_z$ and $\delta B_z$, 
namely, a KAW. A clearer identification of this 
perturbation as a KAW will be given in the 
following. 

\begin{figure}
\plotone{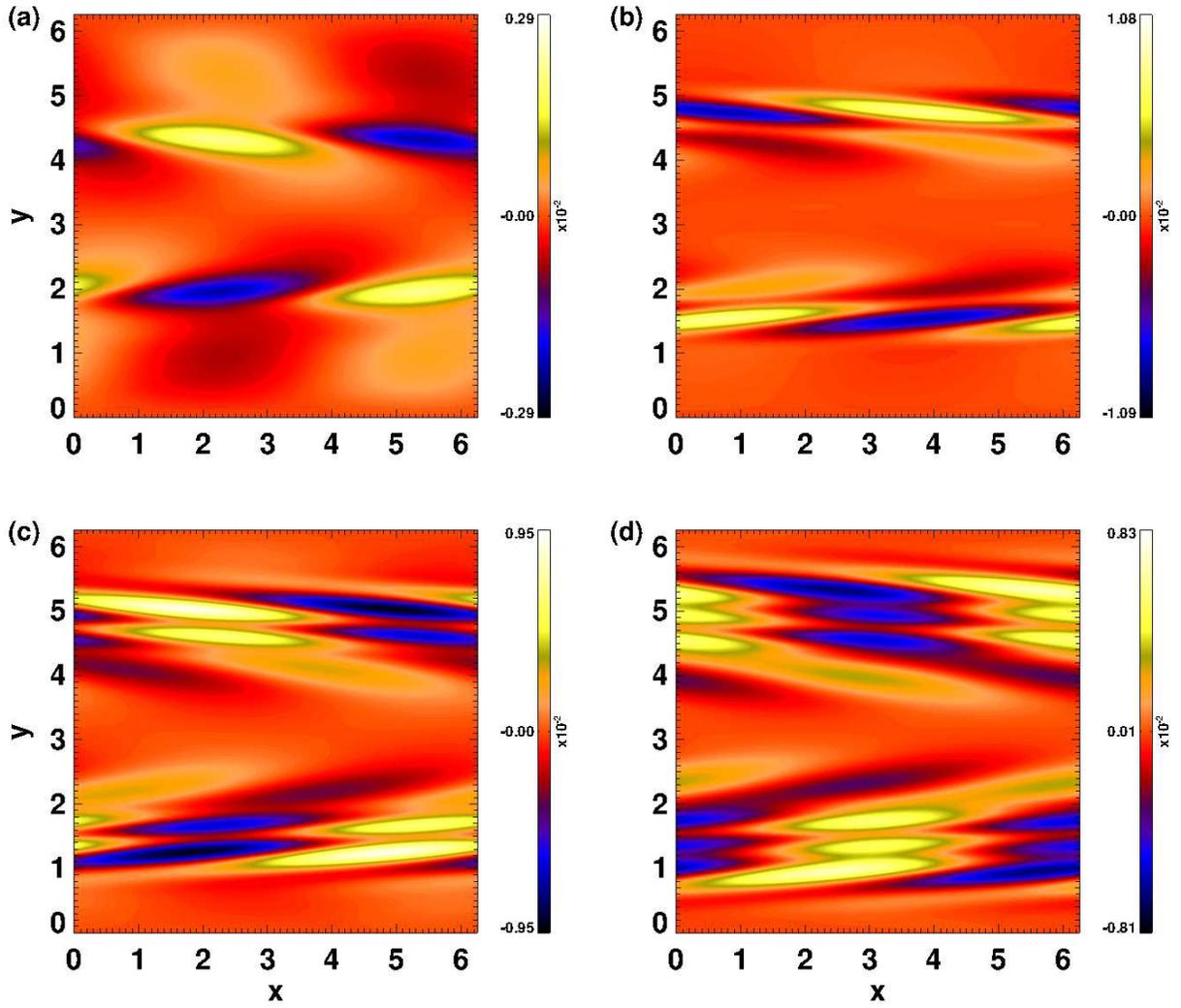}
\caption{The $\delta B_x$ component is plotted in the $xy$-plane at four different times 
for the low-amplitude HMHD run (RUN 2). $t=2$ (a), $t=10$ (b), $t=15$ (c), $t=20$ (d). \label{Fig:expansion}}
\end{figure}

In Section 2.1 we found that the perpendicular group velocity of KAWs is nonvanishing, though 
it is much smaller than the background Alfv\'en velocity. Then, in the 
present configuration, KAWs generated inside the shear layers, while propagating along 
${\bf B}^{(0)}$, slowly drift in the $y$ direction. The perpendicular group velocity 
$v_{g\perp}$ of KAWs has the same sign as $k_y=k_\perp$ (see Fig. \ref{Fig:gvperp}), 
the latter being negative (positive) in 
the shear layer located at $y\simeq 1.8$ ($y\simeq 4.5$). Then, KAWs produced in both shear layers 
would move outside toward the lateral higher-$\beta^{(0)}$ uniform regions. Indeed, examining the 
time behavior of the perturbation, oblique wavefronts progressively occupying the two lateral regions are found. This behavior can be seen in Fig. \ref{Fig:expansion} where the 
$\delta B_x$ component is plotted at four different times. 
We used the following procedure to measure the propagation of these structures in 
the direction perpendicular to ${\bf B}^{(0)}$.

Once the system undergoes phase-mixing, and once wavepackets are generated (at about $t\sim 3$),
we identify all the local maxima and minima of the density field $\rho$ (different fields give similar results).
In practice, following \citet{DonatoEA12}, all the critical
points ${\bf x}_*$ where ${\bf \nabla}\rho=0$ have been found. In these points we computed the
square Hessian matrix of $\rho$, identifying the strongest maxima and minima.
We followed 
the trajectories of these points in time (which are almost parallel each other), that go
from the shear layer out to the border. Taking the average position between the maximum and the minimum,
we calculated its velocity, obtaining $u_y=\simeq \pm 5.4\times 10^{-2}$. 
We can also estimate 
their wavevector components $k_y=\pm 2\pi/\lambda_y\simeq \pm 8.11$ and $k_x=1$. Using these values 
in the expression (\ref{vgparperp}) we obtain for the perpendicular group velocity of the Alfv\'en 
branch (see Fig.s \ref{Fig:gvperp} and \ref{Fig:alfbeta}) the value $v_{g\perp}=\pm 7.1\times 10^{-2}$ for 
$\beta^{(0)}=2.08$ (in the lateral homogeneous region), in reasonable agreement 
with the value  $u_y$ estimated in the simulation. We conclude that the observed structures propagate with 
the group velocity of KAWs. We note also that the propagation angle is $\theta=\arctan |k_y/k_x|\simeq 83^o$, 
close to $\pi/2$ as required for KAWs. 

From Fig. \ref{Fig:gvperp} we note that $v_{g\perp}$ is opposite to $k_\perp$ 
for waves belonging to the SM branch. This implies that SM waves possibly produced 
in the shear layers would laterally drift opposite to KAWs. Thus, the waves observed in the lateral 
homogeneous region cannot belong to the SM mode. Finally, Fig. \ref{Fig:gvperp} 
indicates that for large propagation 
angles the group velocity of FM waves is much larger than that of KAWs, being $\simeq 1.8$ 
for the above values of $k_x$ and $k_y$. This
value is much larger than the lateral propagation velocity $u_y$ of the observed perturbation. 
Consequently, the fluctuation produced by phase-mixing cannot belong to the FM mode.

Another feature we took into account to identify the observed perturbation 
is the polarization. Since the wavevector of these perturbations is nearly parallel to the $y$ direction, 
the condition $\nabla \cdot {\bf B}=0$ implies that the dominant components are $B_{1x}$ and $B_{1z}$. 
We considered these components in the polarization analysis. 
Equations (\ref{B1x}) and (\ref{B1z}) indicate that the magnetic perturbation is 
elliptically polarized. In particular, using the dispersion relation (\ref{disprel}), we 
calculated the quantity
\begin{equation}\label{sigma}
\sigma ({\bf k},\beta_0)= 1-\frac{\omega^2}{c_{A0}^2 k_{||}^2}
\end{equation}
appearing in equations (\ref{B1x}) and (\ref{B1z}). In Fig. \ref{Fig:polar} we report $\sigma$ 
as a function of $k_\perp$ for the three modes, for $k_{||}=1$ and for two values of $\beta_0$, 
corresponding to the lateral uniform region ($\beta_0=2.08$) and to the middle of the 
shear layer ($\beta_0=1.25$). 
From Fig. \ref{Fig:polar} we see that $\sigma$ is negative for Alfv\'en 
and FM waves, while is positive for SM waves. 
Then, for positive $k_{||}$ and $k_\perp$ and given values of $t$ and $x$, 
as one increases the $y$ coordinate, 
the perturbation magnetic field rotates 
clockwise (counterclockwise) in the $zx$ plane 
for Alfv\'en and FM waves (SM waves). 
This characteristic behavior of the eigenmodes
has been compared with the simulation results. 

\begin{figure}
\plotone{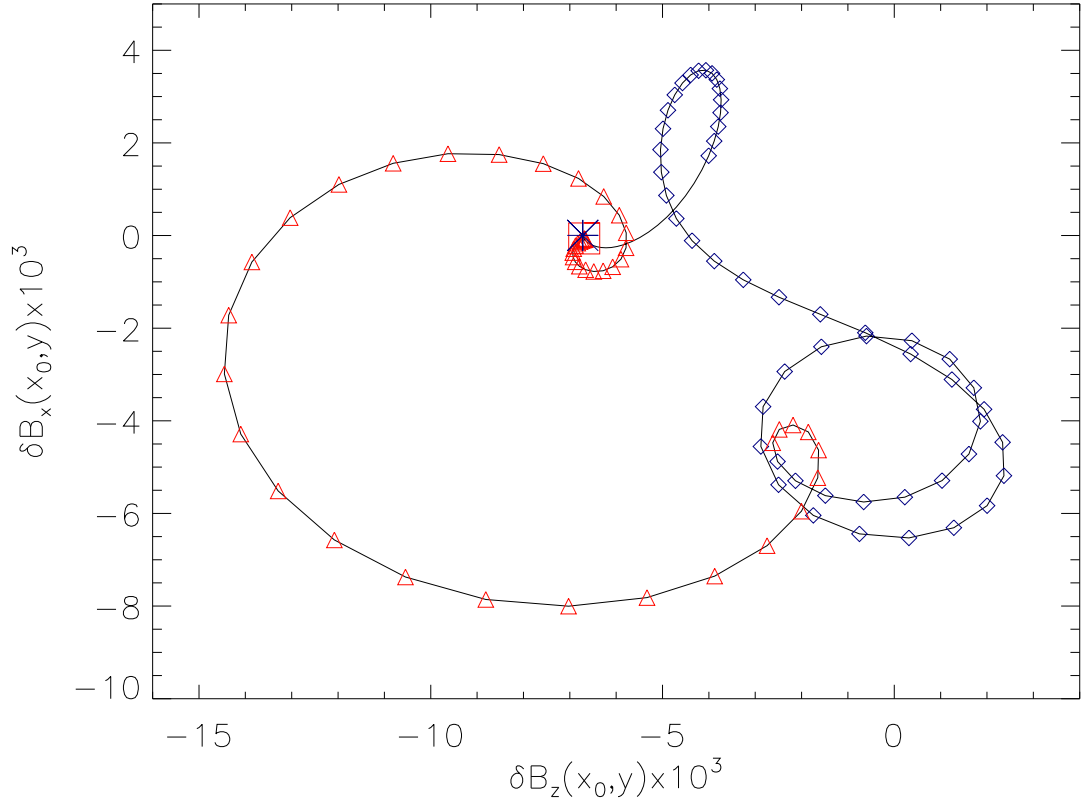}
\caption{The $\delta B_z$ and $\delta B_x$ components are plotted on the two axes 
for $y$ varying in the range $\pi \le y \le 2\pi$, at $x=x_0=\pi$ and $t=20$. 
The central point $y=\pi$ (blue asterisk), the boundary $y=2\pi$ (red square), the shear layer
region (blue diamonds), and the lateral homogeneous regions (red triangles) are indicated.
\label{Fig:hodoHMHD}}
\end{figure}

In Fig. \ref{Fig:hodoHMHD} we plot a hodogram 
in the $\delta B_x$- $\delta B_z$ plane, 
parameterized by the coordinate $y$ 
which varies in the range 
$\lbrack \pi, 2\pi \rbrack$. 
This sample is for a fixed value of the $x$-coordinate
$x=\pi$, and at the time $t=20$. 
At that time, the 
perturbation generated inside the shear layer almost fill the lateral uniform
region.
In the hodogram the blue diamonds indicate the shear layer 
($4 \le y \le 5$) while red triangles indicate the 
lateral uniform region ($5 \le y \le 2\pi$). The
blue asterisk indicates the central point $y=\pi$ and the red square indicates the boundary $y=2\pi$.
It is seen that in most of the shear layer and in the 
lateral homogeneous region the perturbed magnetic field clockwise turns with increasing $y$. Since 
in this region $k_y$ is positive, the observed polarization in the simulation is in accordance 
with that of the Alfv\'en branch.

In conclusion, based both on the group velocity and on the 
polarization analysis,
we deduce that the fluctuations generated inside the shear layers are KAWs. 
From the hodogram we also notice that the perturbation amplitude in the shear layer is 
smaller than in the lateral uniform region. 
Then, the process of KAW generation and their
subsequent lateral propagation tends to move the initial Alfv\'en wave energy away from the shear regions. 
A process of local fluctuating energy depletion is eventually found also in the pure MHD case, but it 
is simply due to dissipation localized 
at the shear layers.

In the hodogram of Fig. \ref{Fig:hodoHMHD} a small perturbation can be seen at the boundary 
between the shear layer and the inner homogeneous region, in which the magnetic field 
counterclockwise turns. This can be interpreted as a SM perturbation propagating from the 
shear layer toward the central region, i.e., opposite to $k_y$, as predicted by the linear theory
(see Fig. \ref{Fig:gvperp}). This SM perturbation can be seen also in Fig. \ref{Fig:expansion}
as a fluctuation drifting toward the center $y=\pi$ of the domain.
However, the energy associated with this perturbation is much lower 
than that of the fluctuation that we have identified as a KAW.

\section{Large-amplitude HMHD and kinetic simulations}
\begin{figure}
\plotone{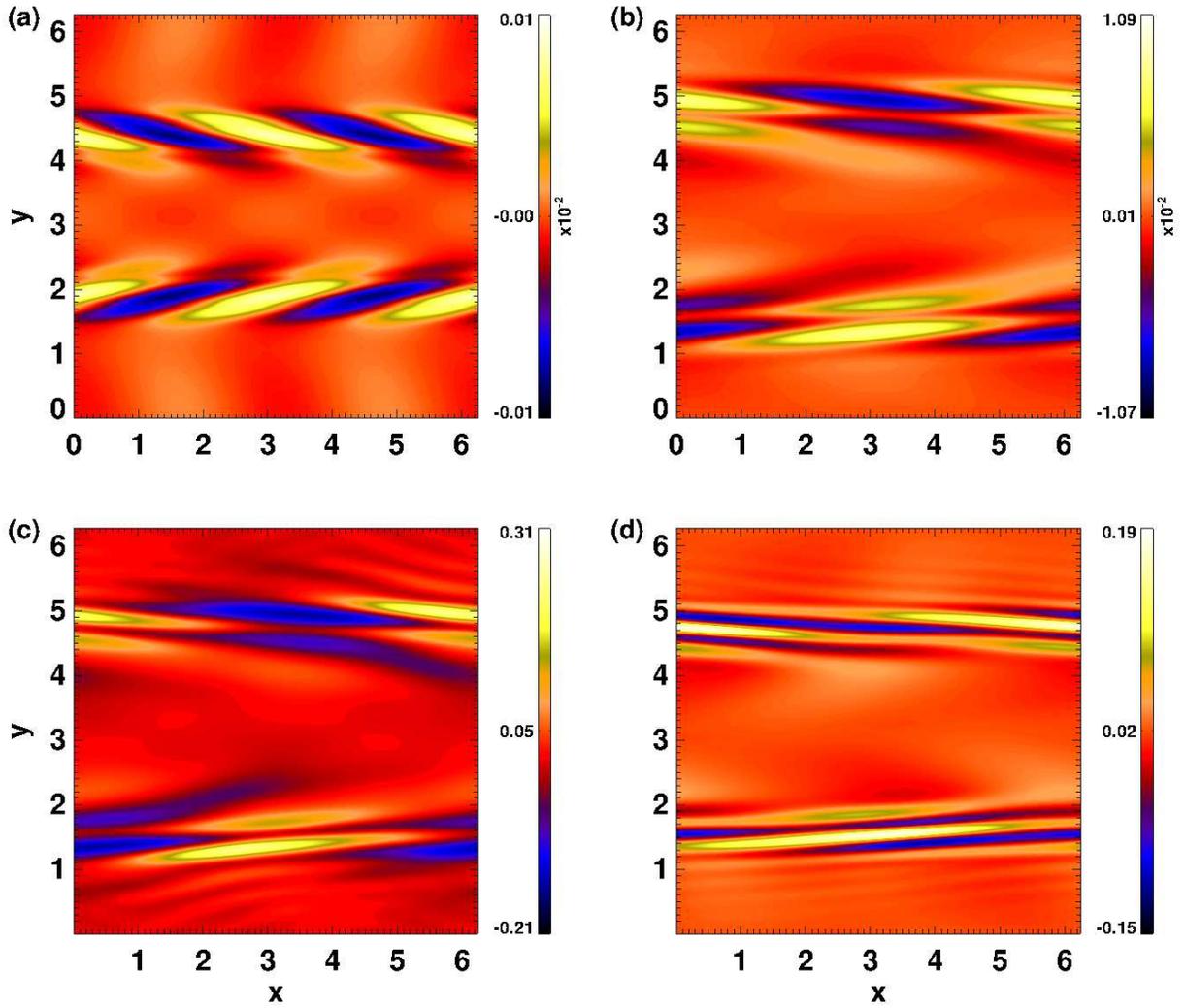}
\caption{The component $\delta B_x$ is plotted as function of $x$ and $y$, at the time $t=13.1$, 
for (a) RUN 1 - MHD with small $a$; 
    (b) RUN 2 - HMHD with small $a$;
    (c) RUN 3 - HMHD, large $a$; and 
    (d) RUN 4 - HVM large $a$. 
\label{Fig:Bxallruns}}
\end{figure}

Next, we consider the phase-mixing of an initial large-amplitude ($a=0.25$) Alfv\'en wave. 
We show results from both a HMHD (RUN 3) and a kinetic (RUN 4) simulation. 
The results are qualitatively similar as in the small-amplitude case, as it can be see in
Fig. \ref{Fig:Bxallruns} where $\delta B_x(x,y)$ is 
illustrated in the $x,y$-plane
for the four runs at time $t=13.1$. 
In particular, also in the large-amplitude case phase-mixing of the initial
wave generates small scale variations 
perpendicular to ${\bf B}^{(0)}$ in the shear layers, mainly in 
the form of KAWs. The identification of these waves 
employs the same 
method as in the low-amplitude case, namely, considering both the 
perpendicular group velocity of perturbations and their polarization.
The main difference between the low and high-amplitude case is that in the latter  
a small-amplitude precursor of the main
KAW perturbation is observed to fill the lateral homogeneous region before the 
arrival of the main perturbation (Fig. \ref{Fig:Bxallruns}). 

The kinetic simulation (RUN 4) has been performed using a HVM
numerical code \citep{valentini07}. The HVM algorithm integrates numerically the
Vlasov equation for the proton distribution function in multi-dimensional phase space. 
In the present work, we restrict our analysis to the 2D-3V (two dimensions in physical space and
three dimensions in velocity space) phase space configuration. The
electrons are considered as a fluid and a generalized Ohm equation is employed for computing
the electric field, which retains the Hall term. In the present work electron inertia effects 
are neglected. Quasi neutrality
is assumed and the displacement current is neglected in the Ampere equation, 
therefore assuming low frequency dynamics. 
Finally, an isothermal equation of state for a 
scalar electron pressure is employed to close the HVM system.
The equations solved by the HVM code are the following:
\begin{equation}\label{vlasov1}
\frac{\partial f}{\partial t}+{\bf u}\cdot \nabla f + \frac{1}{{\tilde \epsilon}}
\left( {\bf E}+{\bf u}\times{\bf B}\right)\cdot \frac{\partial f}{\partial {\bf u}}=0
\end{equation}
\begin{equation}\label{vlasov2}
{\bf E}= -{\bf v}\times {\bf B} +
\frac{{\tilde \epsilon}}{n} \left({\bf j}\times{\bf B}-\frac{{\tilde \beta}}{2}\nabla P_e\right)
\end{equation}
\begin{equation}\label{vlasov3}
\frac{\partial {\bf B}}{\partial t}=-\nabla \times {\bf E} \;\;\; ; \;\;\;
\nabla \times {\bf B}={\bf j}
\end{equation}
where $f(x,y,u_x,u_y,u_z,t)$ is the proton distribution function in phase space 
and ${\bf E}(x,y,t)$ is the electric field. 
The proton density 
$n$ and the ion bulk velocity ${\bf v}$ are obtained as velocity moments of $f$. 
The scalar electron pressure $P_e$ is derived from an isothermal equation of state 
assuming that the electron temperature is equal to the initial (uniform) proton temperature.
In equations (\ref{vlasov1})-(\ref{vlasov3}) all quantities are dimensionless, as 
specified for the HMHD equations (\ref{HMHD1})-(\ref{HMHD4}); moreover, the velocity ${\bf u}$ 
is normalized to ${\tilde c}_A$, the density $n$ to ${\tilde \rho}/m_p$, 
the electric field ${\bf E}$ to ${\tilde c}_A {\tilde B}/c$, 
and the current density ${\bf j}$ to $c {\tilde B}/(4\pi {\tilde L})$.
A detailed description of the numerical method employed to solve equations (\ref{vlasov1})-(\ref{vlasov3}) 
can be found in \citet{valentini07}. The 2D-3V phase space has been discretized 
(see table 1) with $n_x \times n_y = 256 \times 1024$ 
grid points in the spatial domain and 
$51^3$ grid points in the velocity domain.

\begin{figure}
\plotone{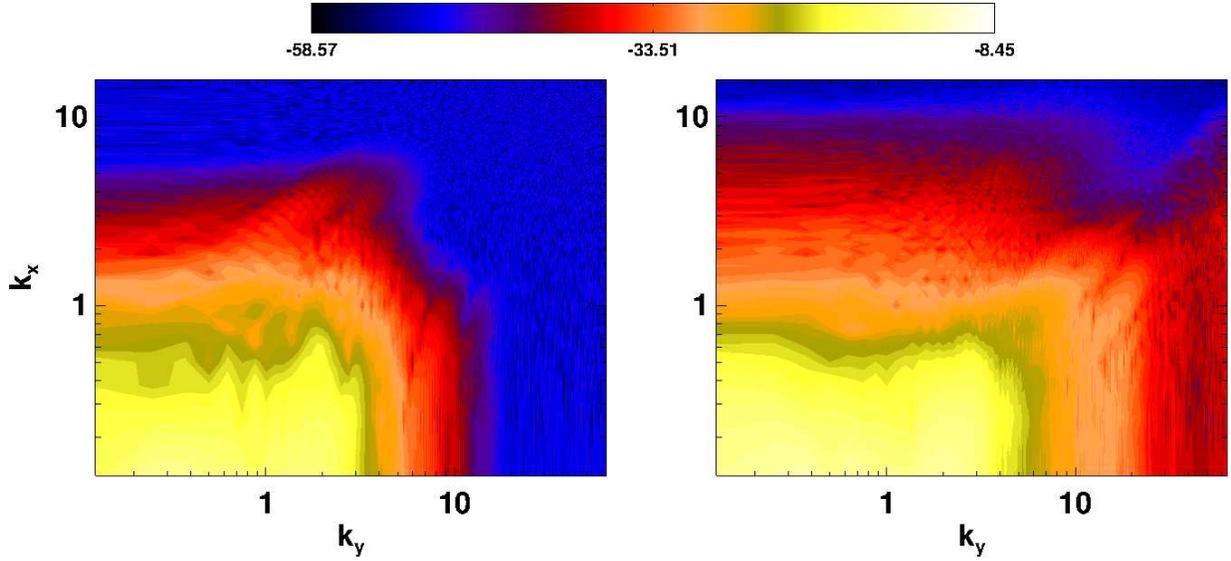} 
\caption{Power spectrum of $\ln{|\delta B|^2}$ from RUN 3 (left panel) and RUN 4 (right panel) at $t=13.1$. \label{Fig:specanis}}
\end{figure}

The initial and boundary conditions (in the physical space) used in RUN 4 are the same as 
those used in the previous large-amplitude HMHD run (RUN 3). 
In the 3D velocity domain, the distribution function $f$ is set equal to zero for 
$|{\bf u}|>u_{max}$, where $u_{max}=5 v_{th,p}$ and $v_{th,p}$ is the proton thermal speed.
Moreover, the initial unperturbed 
proton distribution function is a Maxwellian with a uniform temperature $T^{(0)}=0.5$.
From Fig. \ref{Fig:Bxallruns}
one can observe that 
the fluctuations $\delta B_x$ obtained from the HVM run (RUN 4) at time $t=13.1$ 
are similar to those seen in the HMHD run (RUN 3): 
KAWs develop in the 
shear layers and slowly drift toward the lateral uniform region. 
This similarity is 
probably related to the fact that both models (HMHD and HVM) 
employ the same form for 
the Ohm's law. In Fig. \ref{Fig:specanis}, we compare the power spectra of 
$\ln{|\delta B|^2}$ from RUN 3 (left panel) and RUN 4 (right panel), at the
time $t=13.1$. The contour plots in this figure display similar features, i.e. 
marked anisotropy along the $k_y$ direction due to the phase-mixing process, 
but we can notice that more energy has transferred 
into small scales in RUN 4 relative to RUN 3.
This suggests 
the presence of enhanced small-scale activity when kinetic 
effects are retained in the description of the plasma dynamics.

\begin{figure}
\plotone{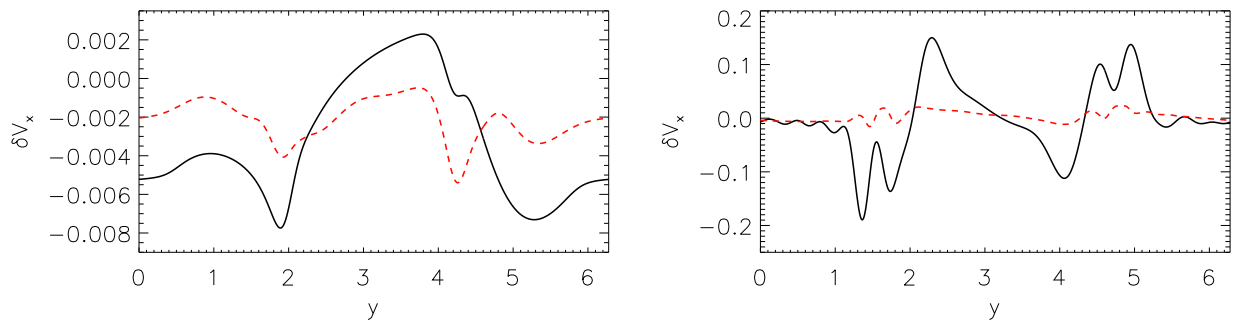}
\caption{$\delta v_x$ profiles are plotted as functions of $y$, for $x=\pi$, at times $t=1$ (left panel) 
and $t=13.1$ (right panel), for RUN 3 (black full line) and RUN 4 (red dashed line).
\label{Fig:Bxdiff}}
\end{figure}

The ion microscopic dynamics which is described by the HVM model 
introduces evidently new effects with respect to the HMHD model,  which are described in the following.
First, we notice that the amplitude of $\delta v_x$ at $t=13.1$ is much lower in RUN 4 than in RUN 3. 
This difference can be better appreciated in Fig. \ref{Fig:Bxdiff}, where profiles of $\delta v_x$ at 
as a function of $y$ at $x=\pi$ and at two different times are plotted. 
It can be seen that the amplitude of $\delta v_x$, which is initially vanishing, increases in time
in the HMHD run, while it remains at a much lower level in the HVM run. Such a different 
behavior is presumably due to kinetic damping 
effects in RUN 4 that act on velocity fluctuations 
parallel to the background magnetic field; such damping 
mechanisms are absent in HMHD. 

\begin{figure}
\plotone{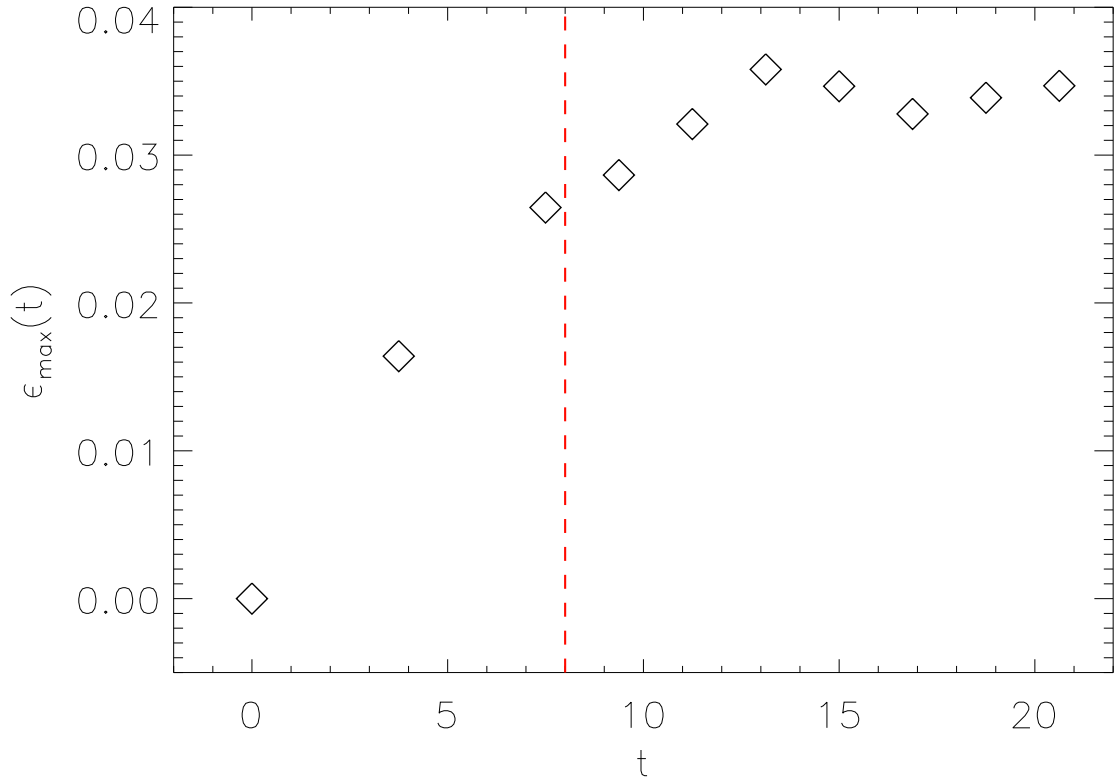}
\caption{The quantity $\varepsilon_{max}$, 
quantifying the departures of the HVM distribution 
function from a Maxwellian, 
is plotted as function of the time. 
The vertical dashed line indicates the time $t=t_d$, described in the text.
\label{Fig:maxeps}}
\end{figure}

\begin{figure}
\plotone{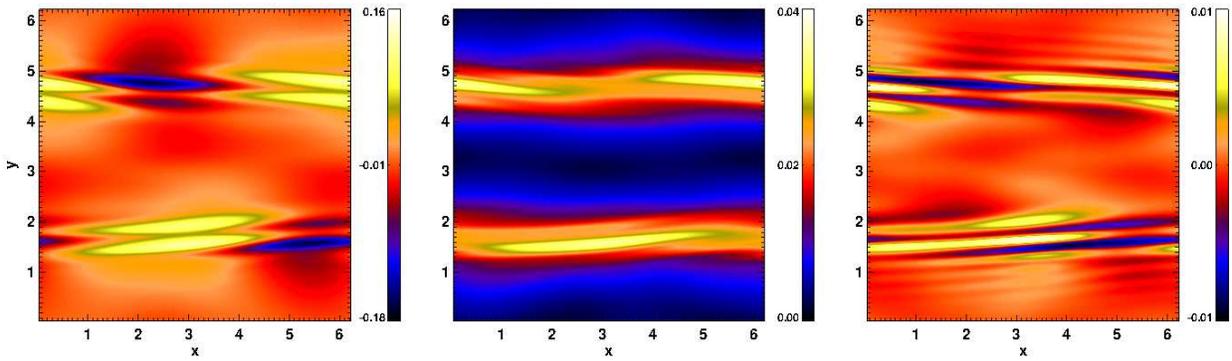}
\caption{The quantities $R$ (left panel), $\varepsilon$ (middle panel), and the electric field parallel 
component $E_{||}$ (right panel) are plotted in the $xy$-plane, 
at time $t=13.1$.
\label{Fig:epsRspace}}
\end{figure}

\begin{figure*}[htpb]
$\begin{array}{ccc}
\epsfxsize=6.5cm \epsffile{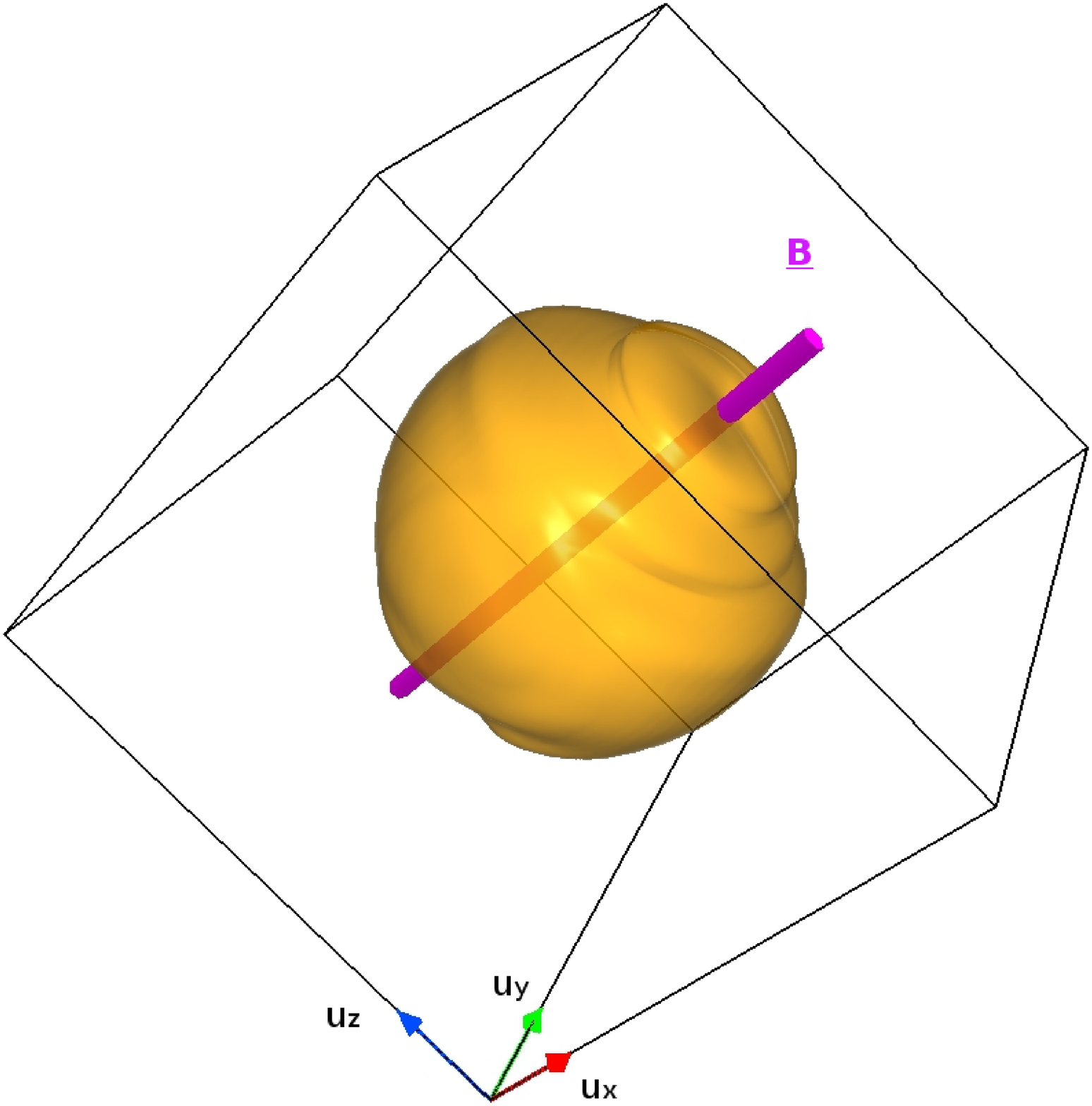} &
\epsfxsize=6.5cm \epsffile{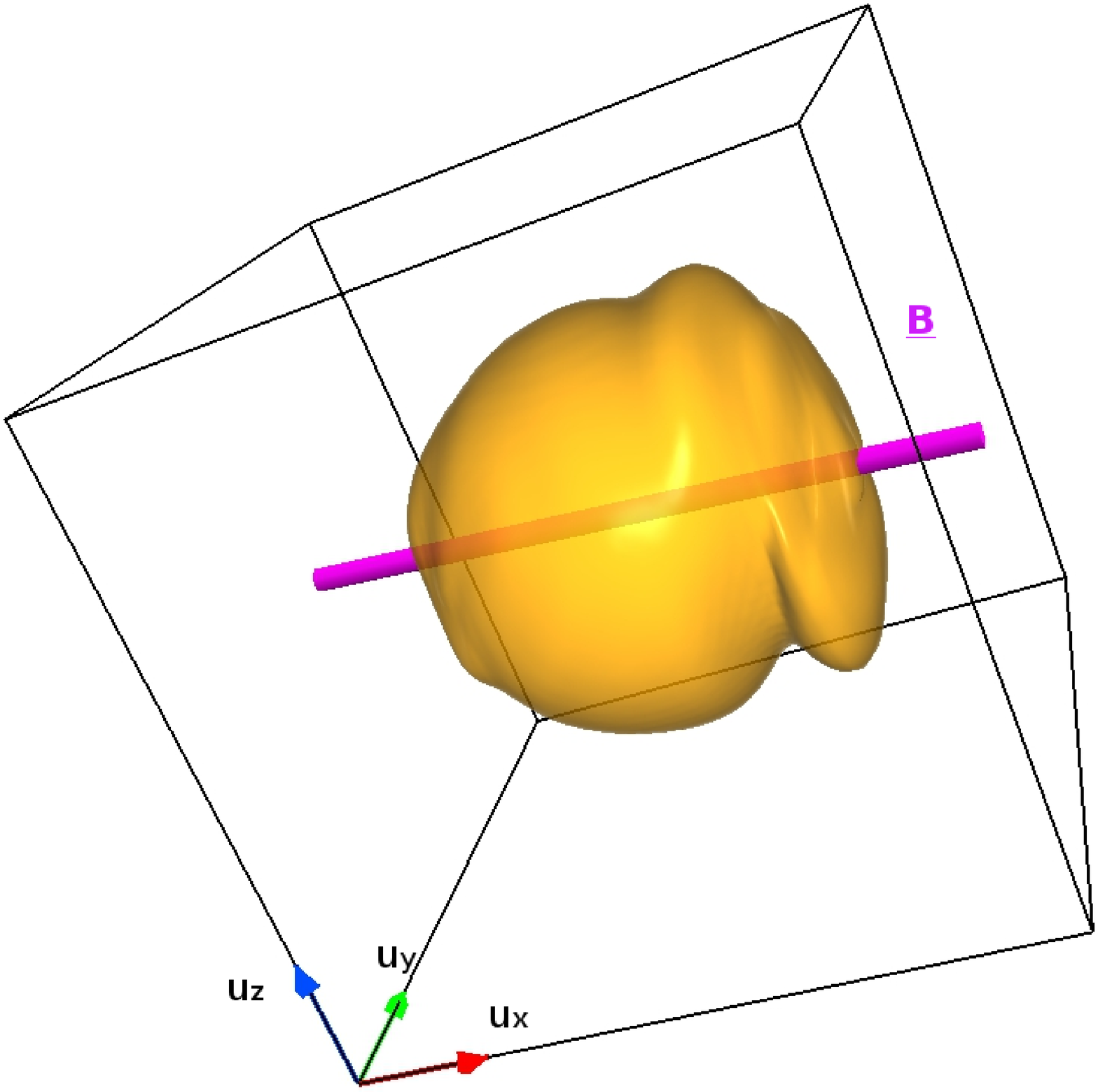} \\
\epsfxsize=8.5cm \epsffile{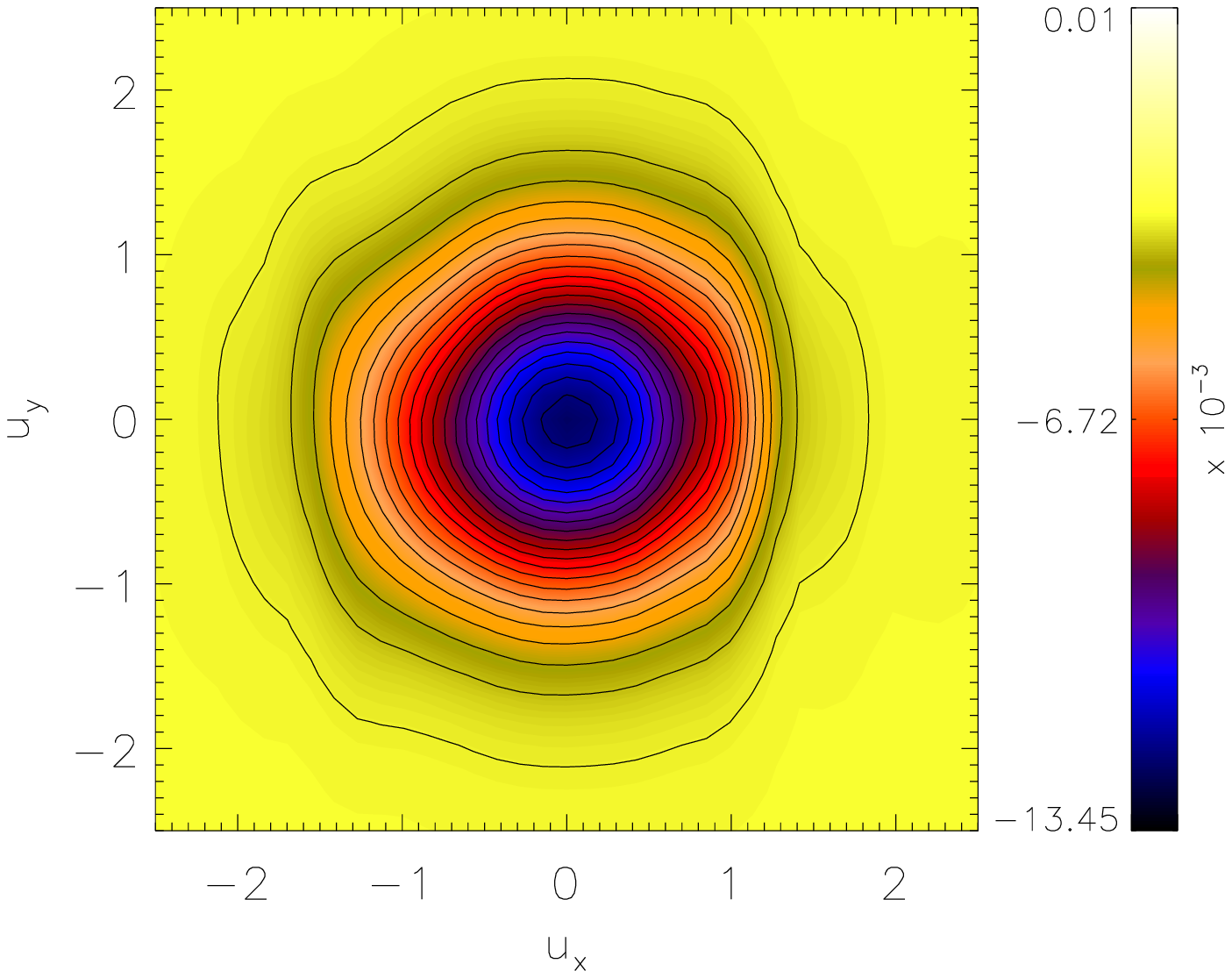} &
\epsfxsize=8.5cm \epsffile{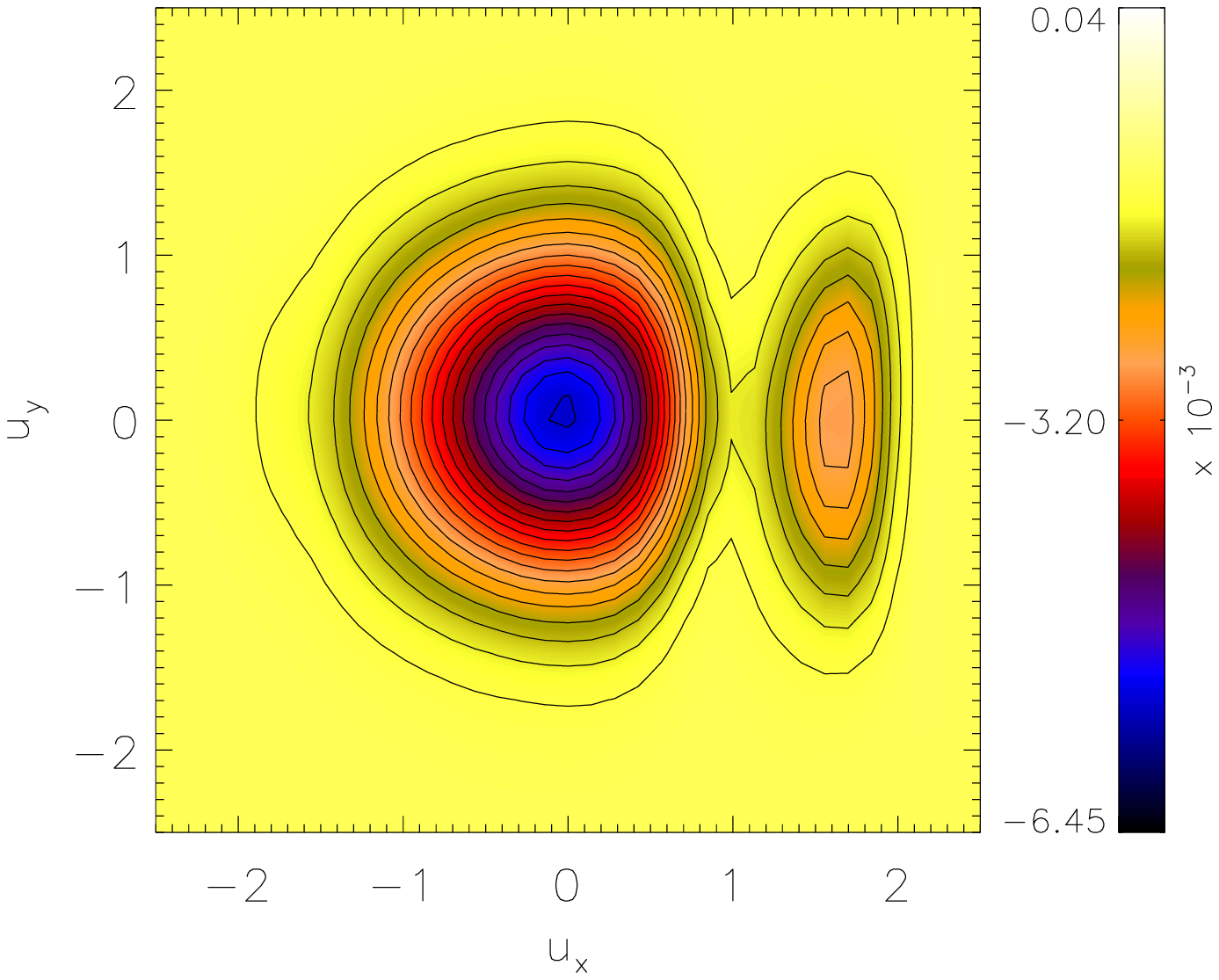} 
\end{array}$
\caption{Top row: surface plot of the proton velocity distribution at the spatial location 
$(x,y)=(6.2,4.7)$ (left), where $\varepsilon$ is maximum and $T_{p\perp}<T_{p||}$, and $(x,y)=(5.4,1.5)$ (right) where
$R$ is minimum, i.e., $T_{p\perp}>T_{p||}$. The magenta tubes in the two plots indicate the direction of the 
local magnetic field. Bottom row: shaded contours (together with level lines) of the proton velocity distribution,
in the $u_x$-$u_y$ plane (at $u_z=0$), in the same spatial locations as in the top row.}
\label{fig:fd} 
\end{figure*}

The HVM models allows us to follow how the proton distribution function is distorted, 
due to resonant interaction of protons with 
the KAW fluctuations, at different space positions and times with respect to the initial Maxwellian (in contrast, 
the HMHD model assumes local thermodynamic equilibrium).
To characterize the departure of the computed 
$f$ from a Maxwellian distribution, we define an
$L^2$-norm difference \citep{greco12, valentini14}:
\begin{equation}\label{epsilon}
\varepsilon(x,y,t)=\frac{1}{n}
\sqrt{ \int \left[f({\bf x},{\bf u},t)-M({\bf x},{\bf u},t) \right]^2 d^3 {\bf u} }
\end{equation}
where $M$ is a Maxwellian with the same density, bulk velocity and isotropic temperature as $f$.
$\varepsilon$ is a positive definite quantity
and may be viewed as a ``distance'' or separation 
between the computed $f$ and an 
equivalent Maxwellian. 
In Fig. \ref{Fig:maxeps} the maximum
\begin{equation}\label{maxeps}
\varepsilon_{max}(t)=\max_{(x,y)\in D} \varepsilon(x,y,t)
\end{equation}
is plotted as a function of time.
We see that $\varepsilon_{max}$ increases in time, eventually saturating at a value $\simeq 0.035$; 
this indicate a progressive departure from a Maxwellian distribution. In Fig. \ref{Fig:maxeps} 
the time $t_d$ is indicated as a vertical red-dashed line, corresponding to the time necessary 
for phase-mixing to produce 
transverse wavevectors comparable with ${\tilde d}_p^{-1}$. At $t\sim t_d$ the growth of 
$\varepsilon_{max}$ becomes slower, indicating that the largest departure from a Maxwellian is 
almost reached when fluctuations at scales of the order of the proton inertial length 
are formed. This is an indication that the KAW fluctuations are responsible for the modifications in the proton distribution function.
Another quantity describing the departure of $f$ from a Maxwellian is the temperature anisotropy
parameter \citep{perrone13}
\begin{equation}\label{R}
R(x,y,t)=1-\frac{T_{p\perp}(x,y,t)}{T_{p||}(x,y,t)}
\end{equation}
where $T_{p\perp}$ and $T_{p||}$ are the proton temperature perpendicular and parallel to the 
local magnetic field ${\bf B}$, respectively; $R<0$ ($R>0$) corresponds to $T_{p\perp}$ larger 
(smaller) than $T_{p||}$. From results of RUN 4 we found that $\varepsilon$ and $|R|$ are clearly 
correlated, the correlation coefficient being $\simeq 0.74$. This indicates that the 
departures of $f$ from a Maxwellian are essentially due to the generation of temperature anisotropy.
In Fig. \ref{Fig:epsRspace} the spatial distributions of $R$ (left) and $\varepsilon$ (middle) are 
plotted at the time $t=13.1$, together with the parallel electric field (right); 
comparing with Fig. \ref{Fig:Bxallruns} we see that the largest departures from a 
Maxwellian are spatially correlated with KAWs, i.e., with fluctuations at transverse scales comparable with $d_p$. 
Positive and negative variations of the anisotropy parameter $R$ with similar amplitudes 
follow one another along the KAW profile. Then, a prevalence of parallel or perpendicular proton kinetic 
energy can equally takes place, due to interactions with KAWs, according to the wave phase. 
A 3D representation (surface plot) of the proton velocity distribution at time $t=13.1$ is shown 
in the top row of Fig. \ref{fig:fd}, for two different spatial locations: $(x,y)=(5.4,1.5)$ (left), 
where $R$ is minimum at the given time, i.e., $T_{p\perp} > T_{p||}$,  and
$(x,y)=(6.2,4.7)$ (right), where $\varepsilon$ is maximum and $T_{p\perp} < T_{p||}$. 
As it is clear from these two plots, 
both distributions depart from a spherical shape, typical of a Maxwellian distribution, and display a structuring 
in the form of rings perpendicular to the local magnetic field, indicating groups of particles 
in resonance with large-amplitude fluctuations.
In the bottom row of the same figure, the contour plots (together with the level lines) of the proton 
velocity distribution in the $u_x$-$u_y$ plane (for $u_z=0$) are presented at the same time and 
spatial locations as for the top row plots.
In particular, the bottom-right panel, corresponding to the location where the ratio 
$T_{p||}/T_{p\perp}$ is particularly large, shows the presence of a well-defined beam of ions 
moving in the direction parallel to the local magnetic field. Such a beam is spatially localized 
where the KAW fluctuations have been generated by phase-mixing, and 
therefore it is reasonable to 
postulate that the beam is produced 
by a wave-particle interaction between protons and the fluctuations 
that we have characterized as KAWs. 

Such interaction could be related to the fluctuating parallel electric field $\delta E_{||}$ 
associated with the KAW. In fact, the Ohm's law (\ref{vlasov2}) allows for the presence of an electric field 
component $E_{||}={\bf E}\cdot {\bf B}/B$ parallel to the magnetic field. This component 
is due to the electron pressure gradient term in equation (\ref{vlasov2}). 
In Fig. \ref{Fig:epsRspace} $E_{||}$ is plotted at time $t=13.1$. It can be seen that 
a fluctuation $\delta E_{||}\sim 10^{-2}$ forms mainly in the shear layers, 
where KAWs are localized. The presence of parallel electric field fluctuations is 
another feature characterizing KAWs. 
In the ideal 
HMHD case $E_{||}$ has no effects on the dynamics 
(the term containing $\nabla P_e/n$ being canceled out when 
calculating $\nabla \times {\bf E}$), unless one includes also a resistivity. 
However, in the HVM model it can have an influence on 
the evolution of the proton distribution function. 
In order to investigate this possibility, we considered 
the potential energy variation (per particle) $\delta U$ 
associated with $\delta E_{||}$, in comparison with the proton thermal energy $E_{th}$. 
The ratio between these two quantities, expressed in our normalized units, 
can be estimated as:
\begin{equation}\label{enratio}
\frac{\delta U}{E_{th}}\sim \frac{8}{3{\tilde \epsilon}{\tilde \beta}} \frac{\delta E_{||}\lambda_{||}}{T_p}
\end{equation}
where $\lambda_{||}$ is the parallel wavelength and $T_p$ is the proton temperature. 
Using the values ${\tilde \epsilon}=0.125$, 
${\tilde \beta}=2$, $\delta E_{||}\sim 10^{-2}$, $\lambda_{||}=2\pi$, and $T_p\sim 0.5$, 
from equation (\ref{enratio}) we find $\delta U/E_{th} \sim 1$. Then, 
the potential energy variation associated with the parallel electric field associated with the KAWs is 
comparable to the proton thermal energy. This implies that $\delta E_{||}$ is 
able to sensibly modify the initial proton distribution function. 
Moreover, the velocity of the proton beam is $u_{beam}\simeq 1.8$. 
We measured the phase velocity of the perturbation along $x$, finding $w_x\simeq 1.6$ which is
comparable with $u_{beam}$. These considerations strongly suggest
that 
the observed field-aligned particle beam is generated by the resonant interaction of protons with
the parallel electric field $\delta E_{||}$ associated with the KAW fluctuations.

\section{Discussion and conclusions}

In this paper we have shown that fluctuations having the character 
of oblique Kinetic Alfv\'en Waves
are readily generated using a simple configuration consisting of an
in-plane two dimensional sheared magnetic field, and an out-of-plane
perturbation that locally propagates as an Alfv\'en wave.
We have described the emergence of the KAWs 
as occurring due to phase-mixing, or refraction of the perturbation 
wave vectors towards angles highly oblique with respect to the 
sheared magnetic field direction. 
Although the refraction is seen in MHD \citep{GhoshEA98},
the basic physical picture of the emergence of KAWs requires
at least a model as complete as HMHD, where it is seen 
at both low and high initial perturbation amplitudes when the 
thickness of perturbation across the magnetic field becomes 
comparable with the proton skin depth. 
For the case of a Hybrid Vlasov solution, at this stage
additional features are observed during this process, 
such as non-Maxwellian proton distributions, temperature anisotropy, 
and the formation a parallel beam in the proton velocity distribution. 
Evidence for resonant wave-particle interaction between 
the beam and the KAW perturbation is also identified. 

The simplicity of the initial setup adopted in these numerical experiments
made it possible to characterize features of KAWs in several ways. These
included, for example, examination of wave polarizations 
and phase speeds based on linear theory.
The identification was facilitated by the choice of the magnetic 
shear as an ideal MHD, HMHD and HVM equilibrium, as well as the choice of 
perturbation as an Alfv\'en mode. However, it is equally 
important to recognize that much of the physical picture described 
in this idealized context is expected to carry
over to more complex configuration, 
including even a full quasi-incompressible turbulence cascade. 
To elaborate on this point briefly we note that 
incompressible couplings, such as those that dominate the 
present models, correspond to quadratic terms, for example 
in the MHD equations (Eqs (2) and (3)) written for constant density.
These nonlinear couplings (we are speaking here of the equations 
prior to separation into equilibrium and perturbation) 
have the familiar property that the Fourier amplitude with wave vector 
$\bf k$ may interact directly with two other wave vectors $\bf p$ and 
$\bf q$ provided that the triadic condition ${\bf k} = {\bf p} + {\bf q}$ is satisfied. 
Suppose we 
identify the sheared magnetic field in 
Eq. (19) as ${\bf B}^{(0)} = (B^{(0)}(y),0,0)$ having 
Fourier modes at wavevectors ${\bf p} \to (0,p_y,0)$.
Now let the perturbation consist of wave vectors 
$\bf q$ and require that it be of an incompressible type but 
otherwise relax  the restrictions of the special case given in Eq. (24).
It is immediately clear that even in a large amplitude situation
more complex than what we treated above, the triadic 
nonlinear couplings will drive excitations to wavevectors
${\bf k} = (0,p_y,0) + (q_x, q_y, q_z)$ that
will acquire increasing values of $k_y$. That is the 
fluctuations will spectrally transfer towards 
wavevectors that are oblique to the sheared magnetic field direction. 
This is closely related to the standard argument
\citep{shebalin83} for perpendicular spectral transfer, 
and is very similar to the nonlinear phase-mixing 
associated with velocity shears and pressure balance structures 
seen in MHD turbulence simulations 
with analogous initial setups \citep{GhoshEA98}.
In this way we can see that the driver of the refraction towards
highly oblique wave vectors is considerably 
more general than in the special case we considered here.
It is reasonable to suppose then, that the Hall and kinetic 
effects that we identified as emerging when the 
transverse scales decrease to the ion inertial length
will also be observed in these more complex circumstances.

We can conclude then that 
the simple mechanism we described here in the context 
of HMHD and Hybrid Vlasov models 
provides a pathway to understand generation of 
fluctuations with the character of Kinetic Alfv\'en Waves.
Driven by magnetic shear, the excitations 
appear at ever smaller scales across the magnetic shear 
layer until the Hall/kinetic effects appear near the ion inertial length
scale. Analysis of polarization and propagation speed 
identifies these fluctuations as being of the KAW-type. 
Particularly interesting is the emergence of a 
field-aligned beam in the self-consistent proton distribution 
function, apparently admitting signatures of 
wave-particle resonances. 
We conjecture that the mechanism described here operates 
also in a quasi-incompressible 
cascade scenario, so that the basic reasoning 
given here may also account for emergence of 
KAW-like features at proton kinetic scales in 
strong turbulence, as suggested based on solar wind observations
\citep{bale05,sahraoui12}.

\acknowledgments
The authors are grateful to P. Veltri for 
many stimulating discussions on the subject of the paper. 
Thanks are due to O. Pezzi for useful suggestions in the analytical 
treatment of the linear problem and to T. Alberti for valuable help with 
simulation data reduction. The numerical HVM simulations have been running on the Fermi 
supercomputer at CINECA (Bologna, Italy), within the ISCRA-C project IsC26-PMKAW.
WHM was partially supported by NASA Grand Challenge 
Research project NNX14AI63G and by NSF AGS-1063439 and AGS-1156094.

\clearpage
%
%
%
%
%
%
%
%
%
%
%
%
%
%
%
%
%
%
%
%
%
%

\clearpage

\begin{table}
\begin{center}
\caption{Simulations setup.\label{tbl}}
\begin{tabular}{c|cccc}
\tableline\tableline
RUN  & Type & Spatial Resolution ($n_x \times n_y$) & Amplitude ($a$)& Hall parameter (${\tilde \epsilon}$) \\
\tableline
1 & MHD & $256 \times 256$ & 0.01 & 0\\
2 & HMHD & $256 \times 256$ & 0.01 & 0.125\\
3 & HMHD & $256 \times 1024$ & 0.25 & 0.125\\
4 & HVM & $256 \times 1024$ & 0.25 & 0.125\\
\tableline
\end{tabular}
\end{center}
\label{tab1}
\end{table}

\end{document}